# On Revolutionizing Power Systems: Suitability of Quantum Computing-Ready Pipelines for Stochastic Unit Commitment Optimization

*Abstract*— **Modern unit commitment (UC) scenarios, particularly those involving uncertainties from demand fluctuations and renewable energy sources, pose significant challenges to traditional optimization methods due to the large search spaces they encompass. This paper proposes a quantum computing-ready optimization pipeline for deterministic and stochastic UC, integrating classical, quantum-inspired, quantum annealing, and hybrid quantum-classical solvers across test systems, including the IEEE 118-bus system. A novel Quadratic Unconstrained Binary Optimization (QUBO) reformulation accommodates generator scaling and 24-hour scheduling horizons with compact Boolean logic penalties for startup, cooldown, and spinning reserve constraints. A divide-and-conquer strategy and Quantum Computing Monte Carlo (QCMC) simulation address scalability across 27,730 uncertainty scenarios, reducing compilation time by up to two orders of magnitude. Experimental results demonstrate the pipeline's ability to achieve efficient, scalable, and accurate UC optimization under uncertainty: the hybrid quantum method approach matches the classical baseline, reduces stochastic UC runtime from 40.76 to 14.79 hours, and characterizes pure quantum annealing as limited to small systems with 0.22% to 2.18% encoding errors and constraint violations. These results provide an analysis of the scalability and computational practicality of hybrid quantum optimization for generator scheduling in UC.**

## NOMENCLATURE

*Indices and Superscripts/subscripts*

| | |
|---|---|
| $a, b, c$ | Indices of a Hamiltonian (Ising/QUBO/PUBO) |
| $d, w, g$ | Demand, wind speed, and solar irradiance uncertainty |
| $i$ | Generator unit |
| $n, m$ | Discrete estimation step |
| $t$ | Time Step (hour) |
| $\kappa$ | Penalty function index |
| $\mathcal{u}$ | Uncertainty scenario |

*Sets*

| | |
|---|---|
| $D$ | Set of generator units |
| $p$ | Set of continuous variables |
| $S$ | Set of uncertainty considerations (i.e., demand, wind speed, solar irradiance) |
| $T$ | Set of total time steps (24 hours) |
| $\tilde{x}$ | Set of binary decision variables |
| $\tilde{\pi}$ | Set of binary-represented continuous decision variables |
| $F^*$ | Set of unit commitment costs for given scenario sets |
| $\mathcal{H}$ | Set of compiled Hamiltonians for given scenario sets |

*Parameters*

| | |
|---|---|
| $A, B, C$ | Fuel cost coefficients for a generator unit |
| $\mathbb{A}$ | Coefficients of binary-represented continuous decision variables |
| $\mathcal{A}, \mathcal{F}$ | Probability and phase estimation unitary |
| $\mathbb{B}$ | Bias terms in a common penalty |
| $G$ | Solar irradiance (W/m$^2$) |
| $G_{norm}$ | Normalized solar irradiance (dimensionless [0,1]) |
| $G_{max}$ | Maximum clear-sky solar irradiance reference (W/m²) |
| $\mathbb{G}, \mathbb{L}$ | Coefficients of binary-represented continuous and binary decision variables for a common penalty |
| $H_{QUBO}$ | A QUBO Hamiltonian |
| $H'_{\mathcal{u}}$ | Modified sub-hamiltonian considering uncertainty for the power balance penalty |
| $h_i$ | Power generation discretization steps for unit $i$ |
| K | Total number of penalty functions |
| $k, c$ | Shape and scale parameters of a Weibull p.d.f. |
| $L_t$ | Demand/load at time t (MW) |
| $\mathbb{M}$ | Coefficients of binary decision variables |
| $N$ | Total number of discrete steps |
| $P_i^{min}$ | Minimum power output for unit $i$ (MW) |
| $P_i^{max}$ | Maximum power output for unit $i$ (MW) |
| $p_t^{wt}$ | Total wind power generation at time $t$ (MW) |
| $p_t^{pv}$ | Total solar PV power generation at time $t$ (MW) |
| $\wp$ | Probability of a function |
| $\mathcal{Q}$ | Grover operator unitary |
| $\mathcal{q}_{ab}$ | Variable interaction strength for a QUBO model |
| $q_i$ | Spinning reserve discretization steps for unit $i$ |
| $u_m^i, d_m^i$ | Minimum uptime and downtime for unit $i$ (hr) |
| $ru_i$ | Ramp-up rate for unit $i$ (MW/hr) |
| $SR$ | Spinning reserve capacity |
| $\mathcal{S}$ | Reflection of a quantum state |

$\mathcal{U}$ Total number of uncertainty scenarios
$v$ Wind speed (m/h)
$\mathcal{Z}, \mathcal{V}$ Rotation unitaries
$\alpha, \beta$ Parameters of the Beta p.d.f.
$\mu, \theta, \phi$ Estimated expectation value, phase shift and global phase of a quantum state
$\mu_s, \sigma_s$ Mean and standard deviation of an uncertainty consideration
$\mu_d, \mu_w, \mu_g$ Means of demand, wind speed, and solar irradiance uncertainty
$\sigma_d, \sigma_w, \sigma_g$ Standard deviations of demand, wind speed, and solar irradiance uncertainty
$\partial_a$ Interactions around a QUBO variable

*Variables*

$F$ Unit commitment cost ($)
$\langle F^* \rangle$ Estimated cost expectation
$p_{t,i,n(m)}$ Power output of a unit $i$ at time step $t$ at estimation step $m$ or $n$. (MW)
$p_{t,i,n}^{sr}$ Spinning reserve of unit $i$ at time step $t$ at estimation step $n$. (MW)
$\lambda_\kappa$ Lagrange multiplier of a penalty $\kappa$
$\mathcal{s}$ Binary slack variable

*Decision Variables*

$z_{t,i,m(n)}$ Binary decision variable for unit $i$ at time step $t$ at estimation step $m$ or $n$.
$x_{t,i}$ Unit commitment variable for unit $i$ at time $t$.
$x_{t,i}^{\uparrow}$ Binary start-up decision variable for unit $i$ at time $t$.
$x_{t,i}^{\downarrow}$ Binary shut-down decision variable for unit $i$ at time $t$.
$y_{ab}$ Decision variables in a QUBO model

*Functions*

$f_d$ Probability density function (p.d.f.) for demand fluctuation
$f_w, f_g$ p.d.f. for wind speed and solar irradiance uncertainty, respectively
$\mathcal{P}_\kappa$ QUBO UC penalty function
$s(t)$ Time-dependent state transition of a quantum system
$\Gamma$ The Gamma function
$MC_q$ Quantum computing Monte Carlo Simulation function
$CV$ Coefficient of Variation

## I. Introduction

UNIT commitment (UC) optimization is a critical economic and social practice in industrial and academic power engineering. The UC problem is also considered an NP-hard problem due to its combinatorial nature, which can be expressed with either a quadratic or linearized objective function [1]. UC objective functions are usually expressed in their quadratic form [2], which imposes increased computational overhead; hence, most UC optimization studies tend to linearize objective functions [3] and grant faster convergence. However, as shown in several readily available solvers, this approach sacrifices information due to the piece-wise sampling of the quadratic cost. Moreover, contemporary UC optimization, both in planning and operations, accounts for risks and uncertainties arising from demand fluctuations and variations in renewable power generation. Two prominent methods for addressing these challenges are robust and stochastic UC (SUC) [4], [5] optimization methods.

Recent advancements in industrial system engineering have sparked a growing interest in quantum computing, which has demonstrated significant improvements in enhancing specific problem formulations, particularly in power engineering. Quantum computing offers numerous advantages in power system engineering, including improved network synchronization using qubit properties [6], enhanced communication security through quantum key distribution [7], and applications in large-scale optimal power flow [8], state estimation [9], and unit commitment [10]–[12]. Given its potential to optimize combinatorial problems, quantum computing holds particular promise for addressing the UC problem, where enhanced computational efficiency and improved value estimation are critical.

Deterministic UC problems formulated as Quadratic Unconstrained Binary Optimization (QUBO) problems can be solved using adiabatic quantum computing. This reformulation, however, results in an NP-complete problem due to the combinatorial complexity of the objective function. Pure

TABLE I
Comparison of Quantum Computing-based Stochastic Unit Commitment Models

| Prior Work | Scale | Constraints | | | | | Scaled Solving | Solution Estimation |
|---|---|---|---|---|---|---|---|---|
| | > 50 generators | Power Balance + Renewable Energy | Startup | Cooldown | Ramping | Spinning Reserve | Separation/ Decomposition Techniques | QAE/ QCMC |
| Kirihara et al. [9] | × | × | × | × | × | × | × | × |
| Nikmehr et al. [10] | × | × | ✓ | ✓ | ✓ | ✓ | Separation | × |
| Hong et al. [11] | × | ✓ | × | × | × | × | Decomposition | ✓ |
| Rotello et al. [12] | × | ✓ | × | × | × | × | × | ✓ |
| Ajagekar and You [13] | × | × | × | × | × | × | × | × |
| Chang et al. [14] | × | × | ✓ | ✓ | ✓ | × | × | × |
| Quinton et al. [15] | × | × | × | × | × | × | × | × |
| Feng et al. [16] | ✓ | × | × | × | ✓ | × | Decomposition | × |
| Mahroo and Kargarian [17] | ✓ | × | × | × | ✓ | × | Separation | × |
| Paterakis [18] | × | × | × | ✓ | ✓ | × | Decomposition | × |
| Proposed Pipeline | ✓ | ✓ | ✓ | ✓ | ✓ | ✓ | Separation | ✓ |

QUBO optimization for the UC problem was demonstrated in [13], [14], in which the UC cost is reformulated in quadratic form, while power balance, security, and other economic constraints are converted into penalty functions. Moreover, the size of the UC problem, measured by the number of decision variables and constraints, grows polynomially with the planning horizon, the number of generating units, and the number of discretization steps. While problem size grows polynomially, the UC problem remains NP-hard in general due to its combinatorial structure [1], [13]. Similar to the findings of [15], hybrid approaches address this challenge by decomposing the UC problem into linear or nonlinear programming components and combinatorial subproblems, enabling distributed optimization techniques to improve computational efficiency [16], [17]. Despite these advancements, significant challenges remain in retrofitting UC for quantum systems, particularly for 24-hour UC problems that require approximately 300 qubits, often exceeding the capabilities of current quantum hardware. Universal quantum computers currently provide up to 1,121 qubits for non-exclusive use, while quantum annealers offer 5,640 qubits. Additionally, UC formulations with timing variables indicate that PUBO exacerbates the qubit requirement when converting to a QUBO model, implying that additional overhead qubits are needed to accommodate continuous variable encodings and the expansion of timing-related variables.

Beyond deterministic UC, quantum research has been extended into distributed and stochastic settings. Nikmehr et al. [10] extended QUBO-based UC to distributed microgrids using quantum-enhanced alternating direction method of multipliers (ADMM), distributing commitment decisions across classical and quantum agents. In the stochastic domain, Rotello et al. [12] demonstrated quantum amplitude estimation to compute expected values in two-stage stochastic programs, thereby establishing a quantum pathway for uncertainty-aware UC. More recent developments have focused on Benders decomposition-based hybrid quantum architecture. Paterakis [18] extended this to a multi-cut Benders approach applied directly to UC using the D-Wave Advantage quantum annealer, exploiting multiple feasible solutions per iteration to accelerate convergence. Hong et al. [11] further demonstrated qubit-efficient quantum annealing for SUC by combining Benders decomposition with quantum ADMM and the augmented Lagrangian method, achieving a qubit requirement independent of the number of Benders cuts. These decomposition strategies represent the current frontier of hybrid quantum UC research and directly inform the extension path of the pipeline proposed in this work.

As shown in Table I, prior studies such as [9]–[18] have explored QUBO reformulations for UC problems, but their scalability evaluations are limited to small systems (e.g., fewer than 10 generator units) and short time horizons (e.g., 6 to 12 hours). These works also omit spinning reserve constraints, which are critical for operational feasibility in real-world applications. Similar to [11], [12], this work presents a two-stage SUC optimization model that incorporates uncertainty variables. However, it further integrates uncertainties in both renewable generation and demand, which is an essential factor for a comprehensive economic assessment. Additionally, this work evaluates the practicality of quantum optimization models by assessing the quality of objective values obtained through quantum annealing and hybrid quantum computing-based optimization. To the best of the authors' knowledge, no prior work presents a unified framework that simultaneously addresses large-scale scalability and spinning reserve modeling for the base UC model. The use of Quantum Computing Monte Carlo (QCMC) simulation, which is based on quantum amplitude estimation, represents another compelling case for a quantum-computing-based pipeline. While prior studies such as [12] and [18] have demonstrated practical applications, this work focuses on exploring the potential advantages and limitations of QCMC in the context of large-scale SUC problems.

This paper addresses these gaps by introducing a quantum-computing-ready SUC pipeline (see Table I) that features a detailed QUBO reformulation incorporating compact binary penalty terms for spinning reserves. The proposed framework is evaluated on systems with up to 54 generators over a 24-hour horizon and includes over 27,000 uncertainty scenarios within a hybrid classical-quantum optimization workflow. A key objective of this work is to characterize the practical limits of pure quantum implementations under current hardware constraints. In particular, pure quantum annealing is restricted to small-scale systems and is susceptible to encoding errors arising from qubit-constrained discretization, which tend to increase with system size. Unlike previous quantum UC approaches that either omit uncertainty or are constrained to limited quantum routines, this work systematically evaluates the suitability of quantum-inspired, annealing-based, and hybrid solver techniques capable of handling realistic UC constraints and large-scale data volumes, thereby analyzing the suitability of quantum optimization including its advantages and limitations for large-scale generator-scheduling UC problems.

Considering the identified gaps in the existing literature, the main contributions of this paper are described as follows:

- A quantum computing–ready optimization pipeline for both deterministic and stochastic UC is proposed to demonstrate its implementation across quantum-inspired, quantum annealing, and hybrid quantum–classical solvers. The proposed framework integrates full-scale classical and quantum-compatible components, enabling end-to-end scheduling on established UC benchmark systems, including the IEEE 118-bus system, under a copperplate network assumption consistent with current quantum solver constraints.
- A novel QUBO reformulation is presented to accommodate generator scaling and 24-hour scheduling horizons, addressing key limitations in earlier QUBO-based UC models. By incorporating compact binary constraints via Boolean-logic penalties, the formulation supports startup and cooldown modeling, similar to classical Mixed-Integer Quadratic Programming (MIQP), while remaining compatible with quantum solvers. Additionally, a structured penalty model for spinning reserve constraints, which is a feature often omitted in prior quantum UC works, is integrated.
- A suitability evaluation of QUBO-based and quantum-compatible optimization methods for solving the UC

problem under uncertainty is presented. By integrating a divide-and-conquer strategy for compiling large stochastic QUBO models and applying QCMC for cost expectation estimation, this paper addresses scalability and sampling efficiency.

This paper is structured as follows: Section II discusses preliminary theories such as quantum computing, quantum optimization, and quantum Monte Carlo. Section III shows the formulation of deterministic and proposed pipelines for stochastic quantum-based UC. Section IV reports the comparative results obtained by classical and quantum deterministic UC. Finally, Section V summarizes the work done in this paper and possible future directions.

## II. Preliminaries

### A. Quantum Computing

Quantum computing, grounded in quantum mechanics, offers a new paradigm for information processing by representing and manipulating quantum states using linear operators. Its effectiveness depends on the hardware, ranging from digital and quantum annealers to analog and universal quantum computers. While universal systems support advanced models, such as adiabatic and topological computing, they are challenging to realize. In contrast, quantum and digital annealers optimize using Hamiltonians, with digital annealers simulating quantum behavior on classical hardware.

### B. Quantum Annealing

AQC adiabatically transitions an optimization problem's initial configuration to the optimal state, mimicking a quantum system's time evolution, as shown in (1).

$$H = s(t)H_i + (1 - s(t))H_f \tag{1}$$

where $H$ is a Hamiltonian corresponds to the current energy state of the quantum system; $s(t)$ represents the state transition as a function of time; $H_i$ denotes the ground state configuration of a quantum system, and the final Hamiltonian $H_f$ is an encoding of an optimization problem.

The adiabatic process shown in Fig. 1 (a) evolves the system by gradually increasing the adiabatic evolution path $s(t)$ with respect to time. Quantum tunneling, as illustrated in Fig. 1(b), is a key quantum mechanical property leveraged in Adiabatic Quantum Computing (AQC) to escape local minima that often trap classical annealing methods.

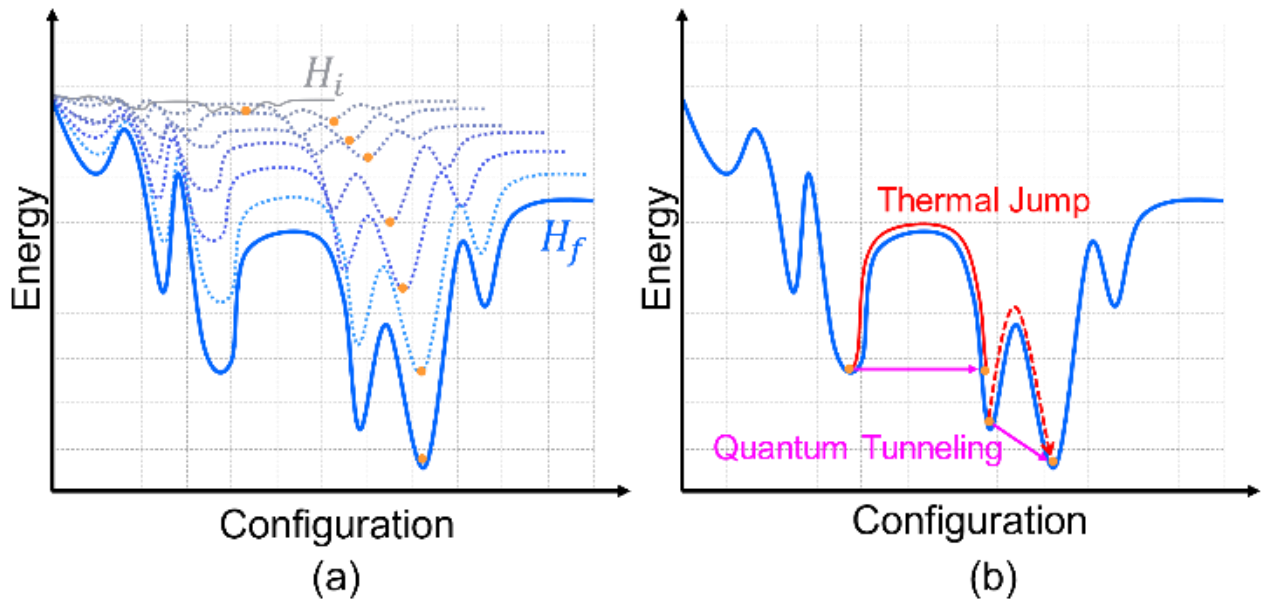

**Fig. 1** Depiction of (a) adiabatic evolution and (b) quantum annealing

### C. Combinatorial Optimization in Quantum Computing

While general quantum advantage has not yet been demonstrated, AQC offers a promising framework for exploring solutions to combinatorial optimization problems. In quantum optimization, problems are modeled so that the system's total energy can be evaluated or minimized. The QUBO model, defined in (2), is structurally equivalent to the Ising model and encodes problems using only binary variables, making it suitable for digital and quantum annealing.

$$\min y^T H_{QUBO} y = \min \sum_{a<b} q_{ab} y_a y_b + \sum_{a} q_a y_a \tag{2}$$

whereas $y$ is a binary variable. The QUBO Hamiltonian $H_{QUBO}$ is a two-dimensional matrix where diagonal terms relate to linear biases $q_a$ and the upper triangular terms are the biases $q_{ab}$ between variables. The equivalence between Ising and QUBO formulations is essential for adapting Hamiltonians to quantum hardware and algorithms. Classical optimization models, which are typically expressed in linear or integer programming, must be reformulated into QUBO or Ising formats, with continuous variables discretized into binary form. However, higher-order terms introduce a polynomial unconstrained binary optimization (PUBO) problem, which complicates QUBO encoding and requires additional reformulation, as illustrated in (3).

$$\min y^T H_{PUBO} y = \min \left( \sum_{a<b<c} q_{abc} y_a y_b y_c + \sum_{a<b} q_{ab} y_a y_b + \sum_{a} q_a y_a \right) \tag{3}$$

PUBO models cannot be directly optimized in quantum annealing-based algorithms since a Hamiltonian representing the optimization problem should only be two-dimensional. As a remedy, a converted PUBO can be reduced to a QUBO [20]. QUBO models can be converted into Ising models, enabling optimization on quantum annealers like D-Wave [21].

### D. Quantum Computing Monte Carlo (QCMC)

Quantum Phase Estimation (QPE) is a foundational quantum computing algorithm for determining a unitary operation's phase, utilizing phase kickback and the Quantum Fourier Transform (QFT). Quantum Amplitude Estimation (QAE) extends QPE to quantify the amplitude of a quantum state $|\Psi\rangle$, incorporating modifications to controlled unitary operators. The estimation of the arbitrary unitary is adapted with a specific emphasis on the Grover operator, referred to as $Q$ herein. Further implementation of the QCMC through (4) is based on the formulation in [22].

$$Q = \mathcal{A} \mathcal{S}_0 \mathcal{A}^\dagger \mathcal{S}_\Psi \tag{4a}$$

$$\mathcal{A} = \sqrt{1-p}|0\rangle + \sqrt{p}\,|1\rangle \tag{4b}$$

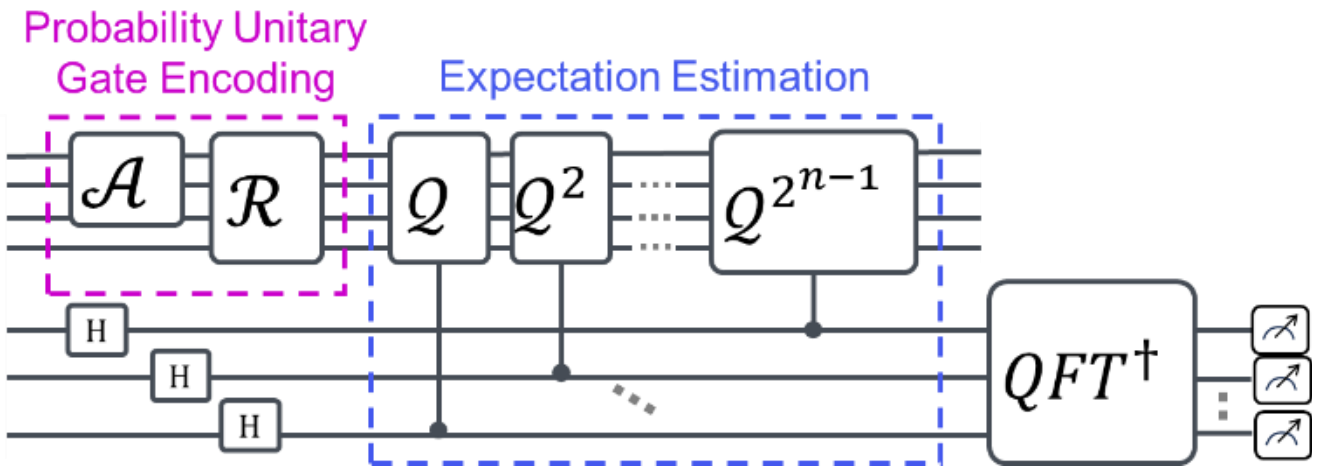

**Fig. 2** Quantum parametrized circuit of quantum computing Monte Carlo (QCMC) simulation.

$\mathcal{S}_0$ and $\mathcal{S}_\psi$ are the reflections of the states $|0\rangle$ and $|\psi\rangle$ respectively. The symbol "†" signifies the conjugate transpose operation. QCMC simulation [22] can be represented as a parametrized quantum circuit, as shown in Fig. 2. A new unitary operation, denoted as $\mathcal{A}$, is defined, where each $\mathcal{p}$ represents amplitudes in a probability density function (p.d.f.). Subsequently, a rotation (5) can be applied to an ancilla qubit $\mathcal{R}$, considering the function $v(x)$ as the binary-to-real mapping function $v(x): \{0,1\}^n \rightarrow \mathbb{R}$.

$$\mathcal{R}|x\rangle|0\rangle = \sqrt{1 - v(x)}|0\rangle + \sqrt{v(x)}\,|1\rangle \quad (5)$$

Such that $Q$ comprises cascaded unitaries (6a), where $\mathcal{F}$ is the qubit phase estimation unitary, while $\mathcal{Z}$ and $\mathcal{V}$ are rotation unitaries. A phase kickback is expected at $Q$. Then an inverse QFT is applied to measure a classical estimate. The quantity to be measured in $Q$ can be obtained from the unitary $\mathcal{V}$. To obtain the measurement $|x\rangle$ from $\mathcal{V}$, (6b) is defined.

$$Q = (\mathcal{F}\mathcal{Z}\mathcal{F}^{\dagger}\mathcal{V})^2 \quad (6a)$$

$$\mathcal{V}|x\rangle = \cos\left(\frac{\theta}{2}\right)|x\rangle + e^{i\phi}\sin\left(\frac{\theta}{2}\right) \quad (6b)$$

$$\mu = \mathbb{E}[v(\mathcal{A})] = \frac{1+\cos\left(\frac{\theta}{2}\right)}{2} \quad (6c)$$

The expectation of a probability unitary $\mathcal{A}$ can be extracted from the real part of $\mathcal{V}$ in (6b) using (6c) where $\mu$ is the estimated expectation value, while $\theta$ is the measure phase of the quantum state. Since QCMC relies on gate-based quantum computing, it cannot run on quantum annealers like D-Wave but can be implemented on hybrid quantum software platforms such as PennyLane.

In the context of SUC, QCMC serves as a sample-efficient alternative to classical MC for estimating the expected UC cost $\langle F^*\rangle$ across large uncertainty sets. Rather than averaging over all scenario costs directly, QCMC encodes the probability distribution of UC costs prob$(F^*)$ into a quantum amplitude estimation circuit, extracting the expectation through phase estimation with quadratic speedup in the number of required samples. As a result, QCMC and classical MC can be viewed as complementary approaches: classical MC offers higher accuracy with sufficient samples, while QCMC reduces the number of oracle queries required to achieve comparable accuracy—an advantage when the uncertainty set is large.

## III. Quantum Computing-Ready Unit Commitment Pipeline

Modern UC optimization requires accounting for numerous operational constraints and uncertainty scenarios, significantly increasing model complexity. In this study, the UC problem is formulated as a quantum-compatible model over a 24-hour horizon under the copperplate network assumption, whereby transmission constraints and power flow equations are neglected, and system-wide power balance is enforced through a single aggregated constraint. In practical system planning, transmission networks are typically designed to satisfy N-1 or N-2 contingency criteria across a range of operating scenarios, which can mitigate the likelihood of severe congestion. However, the copperplate assumption may still lead to infeasible schedules when transmission constraints become binding, particularly in systems with congested corridors or high penetration of distributed renewable generation. Ramping constraints remain physically meaningful under this assumption, as they capture the temporal operating limits of individual generators independently of network topology and line flow constraints. The omission of network constraints does not affect the methodological development of the proposed framework, as the primary focus of this work is on the design of a quantum computing pipeline for SUC. Nevertheless, incorporating network constraints represents an important direction for future research.

To address the computational challenges of large-scale SUC, this paper integrates QCMC techniques for sample-efficient expectation estimation. It presents a divide-and-conquer strategy to streamline the compilation of QUBO Hamiltonians. As shown in Fig. 3, the proposed hybrid quantum pipeline begins with classical preprocessing steps, such as forecasting and uncertainty modeling, followed by UC formulation and transformation into QUBO or Constrained Quadratic Mode (CQM). For stochastic scenarios, the divide-and-conquer method decomposes each scenario's QUBO model into smaller sub-Hamiltonians, which are compiled independently and then aggregated. This modular strategy ensures efficient compilation while preserving constraint structure. The pipeline distributes optimization workloads across classical and quantum resources, leveraging central processing units (CPU), graphics processing units (GPU), and Quantum Processing Units (QPU) to solve deterministic and stochastic UC problems efficiently. This synergistic integration of classical and quantum methods enables scalable, uncertainty-aware UC optimization for generator-scheduling problems, forming a modular foundation extensible to network-constrained settings via Benders decomposition.

### A. Deterministic UC Problem QUBO

The proposed QUBO formulation for the UC problem incorporates a multi-time-step plan for managing multiple generators in power grids. The formulation retains a quadratic objective function while converting continuous variables into binary representations.

**Objective Hamiltonian.** This paper derives QUBO Hamiltonians from a generic MIQP-based UC objective function and standard UC constraints (see Appendix A). The proposed generic QUBO-based UC formulation comprises six Hamiltonians, including an objective function and five penalty terms.

$$\min F = \min y^T H_{PUBO} y = \min y^T \left[H_{obj} + \sum_{\kappa \in K} \lambda_\kappa \mathcal{P}_\kappa\right] y \quad (7)$$

Where $H_{QUBO}$ is the overall QUBO UC Hamiltonian, $H_{obj}$ is the objective Hamiltonian, and $\mathcal{P}_\kappa$ is a vector representation of the penalty Hamiltonians wherein $\lambda_\kappa$ is a vector representation of the Lagrange multiplier of each penalty. The computed cost for the unit commitment problem is represented as $F$.

$$H_{obj} = H_{obj}^{lin} + H_{obj}^{quad} \quad (8)$$

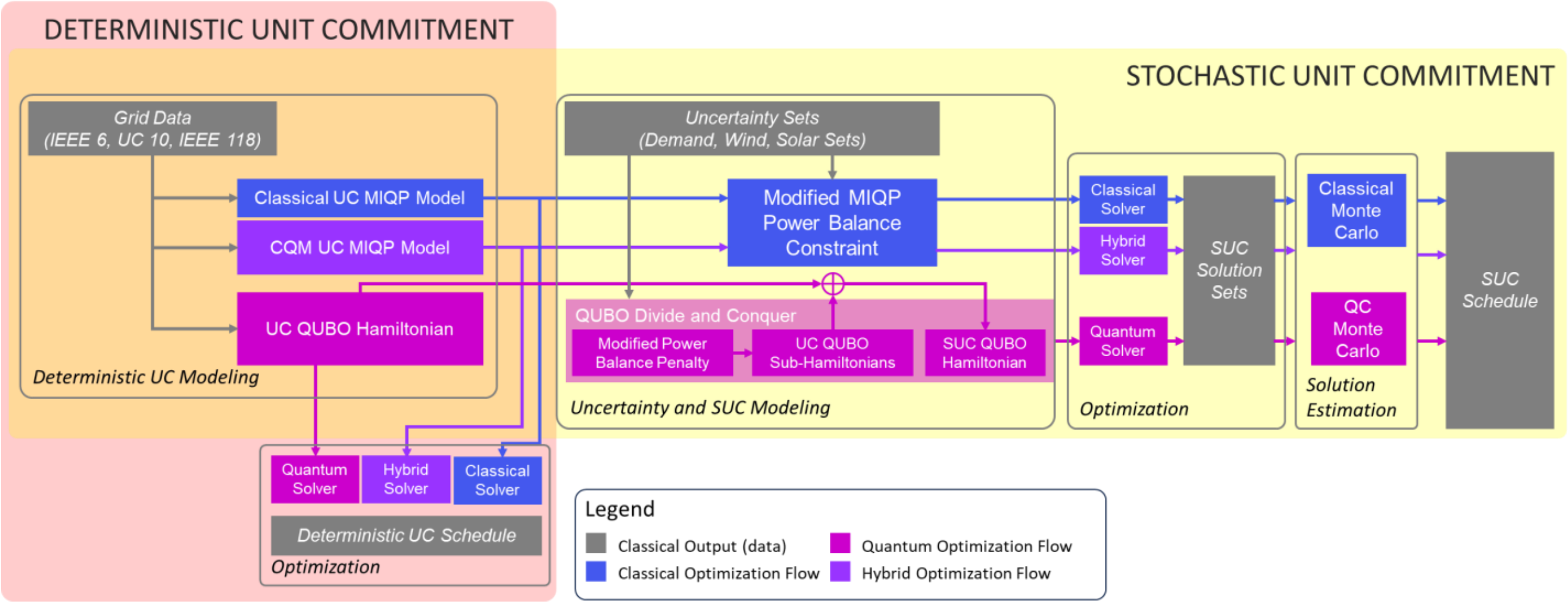

**Fig. 3** Proposed pipeline for the stochastic UC with classical and quantum modelling and optimization options starting from the deterministic modeling to stochastic UC estimation.

where $H_{obj}^{lin}$ are the linear terms of $H_{obj}$ corresponding to the quadratic fuel cost and $H_{obj}^{quad}$ corresponds to the interaction of each binary variable.

**Penalties.** The constraints of a classical MIQP model are reformulated as penalties $\mathcal{P}$. The penalties for the QUBO model are as follows:

$$\lambda_a\left((\tilde{\pi}^T\mathbb{A}\tilde{\pi}+\mathcal{s}_a)^T(\tilde{\pi}^T\mathbb{A}\tilde{\pi}+\mathcal{s}_a)\right)^2 \tag{9a}$$

$$\lambda_b\left((\tilde{x}^T\mathbb{M}\tilde{x}+\mathcal{s}_b)^T(\tilde{x}^T\mathbb{M}\tilde{x}+\mathcal{s}_b)\right)^2 \tag{9b}$$

$$\lambda_c\left(\begin{pmatrix}(\tilde{x}^T\mathbb{G}\tilde{x}+\mathcal{s}_{c_1})^T(\tilde{x}^T\mathbb{G}\tilde{x}+\mathcal{s}_{c_1})+\\(\tilde{\pi}^T\mathbb{L}\tilde{\pi}+\mathcal{s}_{c_2})^T(\tilde{\pi}^T\mathbb{L}\tilde{\pi}+\mathcal{s}_{c_2})\end{pmatrix}-\mathbb{B}\right)^2 \tag{9c}$$

(9a) represents the reformulated quadratic power balance penalty, where $\mathbb{A}$ is the coefficient matrix and $\tilde{\pi}$ denotes the binary-encoded vector representing continuous power generation. Although the expression $\tilde{\pi}^T\mathbb{A}\tilde{\pi}$ implies higher-order interactions due to binary encoding, a symbolic PUBO-to-QUBO reduction is applied to ensure all resulting terms are at most quadratic, maintaining QUBO compliance. Equation (9b) captures the penalties for binary decision variables $\tilde{x}$ related to uptime, downtime, and ramping constraints as $\mathbb{M}$ denotes the corresponding coefficients. Eq. (9c) addresses penalties involving both power generation and binary decision variables, such as spinning reserves, with $\mathbb{G}$ and $\mathbb{L}$ as coefficients for binary and continuous variables, respectively, and $\mathbb{B}$ representing bias terms. Conversion of inequalities into equalities include slack variables $\mathcal{s}$. Detailed reformulations are be found in Appendix B. It is also noted that the reduction is performed using symbolic transformations via PyQUBO [23], which ensures that all higher-order binary terms are reduced to pairwise interactions suitable for QUBO solvers.

**Compact Logical Penalties.** Penalties with pure binary constraints with 0/1 bounds, such as uptime and downtime constraints, can be compressed using Boolean logic. Multiple binary variables would result in complicated reformulations. However, using the Python package PyQUBO simplifies encoding such complex logical penalties. The uptime constraint is expressed as (10a) given the binary startup-decision variable $x_{t,i}^{up}$ and the binary UC variable $x$.

$$1 \geq x_{t,i}^{\uparrow} + x_{t,i} - x_{t-1,i} \geq 0 \qquad \forall\, t \in T, i \in D \tag{10a}$$

$$\left(x_{t,i}^{\uparrow} \oplus x_{t,i}\right) \vee \neg\left(x_{t,i} \oplus x_{t-1,i}\right) \tag{10b}$$

Equation (10b) presents a simplified approach for encoding classical 0/1-bounded uptime and downtime constraints as logical QUBO penalties. In this formulation, the symbol "$\oplus$" denotes a logical exclusive OR, " $\vee$ " indicates a logical inclusive OR, and "$\neg$" represents a logical NOT operation The variables $x_{t,i}^{\uparrow}$ and $x_{t,i}$ are binary decision variables corresponding to start-up and unit commitment status, respectively.

Boolean compaction is applicable to constraints in which all decision variables are binary with 0/1 bounds and the constraint can be expressed as a logical proposition. In the proposed QUBO formulation, four types of constraints satisfy this condition: generator startup, shutdown, minimum uptime, and minimum downtime constraints in (21)–(26), corresponding to penalties $P_3 - P_8$ in Appendix B. Constraints involving continuous variables, including power balance (19), generation limits (20), and spinning reserve constraints (27)–(29), are not suitable for Boolean compaction and instead require the introduction of explicit slack variables followed by PUBO-to-QUBO reduction. The variable reduction achieved through Boolean compaction becomes significant at scale. For a system with $|D|$ generators over $|T|$ time steps, each compacted constraint type eliminates $O(|D||T|)$ slack and auxiliary variables. Across the four applicable constraint types, the total reduction is approximately $4|D||T|$ corresponding to savings of 72, 1,056, and 5,184 variables for the IEEE 6-bus, 11-generator, and IEEE 118-bus systems, respectively.

## *B. Uncertainty Modeling and Scenario Generation*

Uncertainty in the unit commitment (UC) model arises from fluctuations in demand and the intermittency of power generation from renewable energy sources (RES). Demand variability is modeled using a normal distribution, as defined in (11).

$$f_d(L,\mu_d,\sigma_d) = \frac{1}{\sigma_d\sqrt{2\pi}}\,exp\left[-\left(\frac{(L-\mu_d)^2}{2\sigma_d^2}\right)\right] \tag{11}$$

where $\sigma_d$ and $\mu_d$ is the standard deviation and mean of the demand profile respectively and $L$ is the demand in each period. Uncertainty of RES is modeled by considering intermittency in the wind speed $v$ (m/s) and solar irradiance $G$ (W/m$^2$) in (12) and (13).

$$f_w(k,c,v) = \frac{k}{c} \cdot \left(\frac{v}{c}\right)^{k-1} e^{-\left(\frac{v}{c}\right)^k} \tag{12}$$

$$f_g(\alpha,\beta,G_{norm}) = \frac{\Gamma(\alpha+\beta)}{\Gamma(\alpha)\Gamma(\beta)} G_{norm}^{\alpha-1}(1-G_{norm})^{\beta-1} \tag{13}$$

The probability distribution of wind speed is modeled using the Weibull distribution, as shown in (12), where $c = 8.2$ and $k = 2$ denote the scale and shape parameters, respectively. The Weibull distribution is commonly applied in wind energy studies due to its empirical accuracy in representing wind speed variability across diverse geographical locations [24]. The distribution in (13) represents solar irradiance uncertainty using the Beta distribution with parameters $\alpha = 26$ and $\beta = 13$. The Beta distribution is appropriate in this context because its bounded support over [0,1] and flexible shape make it well-suited for capturing the normalized variability of solar insolation [25]. Solar irradiance $G$ (W/m²) is normalized as $G_{norm} = G/G_{max} \in [0,1]$ using a clear-sky reference of $G_{max} = 1000$W/m², consistent with the bounded support of the Beta distribution. The parameters $\alpha = 26$ and $\beta = 13$ characterize the shape of the normalized distribution, with $G_{norm}$ scaled back to physical units for integration into the power balance constraint (15b). Considering the combined uncertainties from demand fluctuations and renewable energy intermittency, the required sample size is determined using (14a).

$$|\mathcal{U}| = \max(\{u(\sigma_s,\mu_s) \mid \forall s \in S\}) \tag{14a}$$

$$u(\sigma_d,\mu_d) = \frac{1}{\varepsilon^2}\left(\frac{\sigma_d}{\mu_d}\right)^2 \qquad \forall s \in S \tag{14b}$$

$$\mu_w = c\Gamma\left(1+\frac{1}{k}\right), \sigma_w = c\sqrt{\Gamma\left(1+\frac{2}{k}\right) - \left(\Gamma\left(1+\frac{1}{k}\right)\right)^2} \tag{14c}$$

$$\mu_g = \frac{\alpha}{\alpha+\beta}, \sigma_g = \sqrt{\frac{\alpha\beta}{(\alpha+\beta)^2(\alpha+\beta+1)}} \tag{14d}$$

$u(\sigma_s,\mu_s)$ is a function computing the sample size for a given probability distribution where $\sigma_s$ and $\mu_s$ is the standard deviation and mean of the distribution, respectively, and $\varepsilon$ is the marginal error at 1%. The subscript $s$ may be $d$, $w$ or $g$ which denote the demand, wind speed or solar irradiance, respectively. The sample size for the joint uncertainties $\mathcal{U}$ is computed as the maximum of the calculated sample sizes for each uncertainty p.d.f. The sample size for the demand scenarios is computed using (14b) with $\mu_d \approx 3.7$ GW and $\sigma_d \approx 0.185$ GW representing the mean and standard deviation of daily demand over the 24-hour horizon, with $\sigma_d$ set to 5% of $\mu_d$. These aggregates are used solely for sample size determination via the dimensionless ratio $\sigma_d/\mu_d$, consistent with the MW-based UC formulation. The resulting 27,730 samples represent hourly demand profiles drawn from the normal distribution (11) at the hourly MW scale. Weibull distribution's standard deviation ($\sigma_w \approx 3.80$) and mean ($\mu_w \approx 7.26$ ) are computed using (14c), resulting in a sample size of 2,732 scenarios. Similarly, using (14d), the Beta distribution yields $\sigma_g \approx 0.07$ and $\mu_g \approx 0.6\overline{7}$, corresponding to a sample size of 125 scenarios.

The joint scenario sample size $|\mathcal{U}|$ is determined as the maximum across the individual uncertainty distributions under the assumption that demand, wind speed, and solar irradiance are mutually independent. Under this independence assumption, the joint distribution factorizes as $f(d,w,g) = f_d \cdot f_w \cdot f_g$, and the sample size required to achieve $\varepsilon$-accuracy in estimating the joint expected cost is bounded by the most conservative individual sample size: the maximum across the three distributions as defined in (14a). This conservative selection ensures adequate marginal coverage for each uncertainty source. It is acknowledged that the independence assumption may not hold in practice. Weather-driven correlations such as positive coupling between solar irradiance and cooling demand, or seasonal wind-demand dependencies can produce joint tail events that are underrepresented in an independent sampling scheme. In such cases, joint distribution would exhibit heavier tails than the product of marginals, potentially requiring a larger scenario set or correlated sampling approaches such as copula-based methods for adequate tail coverage. Incorporating correlation structures into the scenario generation framework is identified as a direction for future work, particularly for systems with strong weather-driven renewable-demand coupling.

### C. Stochastic Optimization Routine with QCMC

The hybrid CQM solver performs a structural separation of the UC problem formulated in (7), rather than relying on iterative decomposition. By classifying terms in $H_{QUBO}$ by type, the CQM framework routes linear objectives and linear constraints from (9b), involving only binary commitment variables $\tilde{x}$, to classical hardware (CPU/GPU), while quadratic terms in (9a) and (9c), involving continuous power and reserve encodings $\tilde{\pi}$, are formulated as QUBO subproblems submitted to the QPU [23]. This single-pass partitioning based on term structure eliminates the need for iterative subproblem exchange and convergence criterion, distinguishing it from decomposition-based methods such as Benders or ADMM. Optimality is governed by the classical branch-and-bound applied to the binary commitment variables, with QPU-derived samples providing guidance rather than solely determining the final solution. The modified power balance penalty $\mathcal{P}_1^*(u)$ for each stochastic scenario $u$ is then introduced as (15a).

The proposed SUC formulation follows a two-stage stochastic programming framework. In Stage 1, the commitment decisions $x_{t,i}, x_{t,i}^{\uparrow}$, and $x_{t,i}^{\downarrow}$ are determined prior to the realization of uncertainty and are fixed across all scenarios, thereby satisfying the non-anticipativity constraint. In Stage 2, the dispatch is optimized for each scenario through the scenario-specific power balance penalty $\mathcal{P}_1^*(u)$, which adjusts the residual demand $p_{u,t}^r$ based on realized renewable generation and demand. This structure is reflected in the proposed divide-and-conquer algorithm: $H^*$ is compiled once to encode the Stage 1 commitment decisions, while $H^{*\prime}$ is constructed for each scenario to capture the corresponding Stage 2 adjustments. The expected cost, denoted as $\langle F^* \rangle$, is defined as the expectation of the Stage 2 dispatch cost over the uncertainty set $\mathcal{U}$, conditioned on the fixed Stage 1 schedule.

$$\mathcal{P}_1^*(\mathcal{u}) = \sum_{t \in T} \left( p_{u,t}^r - \sum_{i \in D} \sum_{n}^{N} p_{t,i,n} z_{t,i,n} \right)^2 \quad (15a)$$

$$p_{u,t}^r = L_t - p_{u,t}^{wt} - p_{u,t}^{pv} \quad (15b)$$

whereas $\mathcal{P}_1^*$ is the modified penalty function of the power balance constraint; $p_t^{wt}$ and $p_t^{pv}$ are the wind and solar PV power generation at time $t$, respectively. The uncertainty sample set from (14a) results in 27,730 samples that would need to be initially modeled for both classical and quantum UC. Fig. 4 depicts the stochastic optimization procedure for different optimization options considering combinations of classical and quantum model optimization and expectation value computation. Classical UC problems are reformulated into QUBO or CQM for optimization using D-Wave's hybrid solver. Uncertainty scenarios are incorporated by recombining power balance constraint Hamiltonians with base QUBO models. Recorded costs from each scenario optimization are analyzed using either classical Monte Carlo or quantum computing Monte Carlo to estimate the cost expectation $\langle F^* \rangle$.

### *D. Divide-and-Conquer Strategy for SUC QUBO Model Recombination*

The compilation time of QUBO models scales with the number of binary variables and their interactions, with most time attributed to compiling the objective Hamiltonian $H_{obj}$. To address this, QUBO Hamiltonians are decomposed into sub-Hamiltonians, allowing for the separate compilation of penalties related to uncertainty. The proposed divide-and-conquer algorithm addresses the scalability challenge of QUBO-based SUC by structuring the solution process into three key phases: divide, conquer, and recombination. The algorithm decomposes the high-dimensional stochastic problem into scenario-specific QUBO subproblems, solves each independently, and then aggregates the results to estimate the expected system cost efficiently:

**Divide Phase.** In this phase, the algorithm partitions the SUC problem into independent QUBO sub-Hamiltonians, each corresponding to a distinct uncertainty scenario. This decomposition reduces memory overhead, shortens compilation time, and ensures compatibility with fixed-size QUBO solvers, all while preserving the essential structure of the UC problem. Each scenario-specific QUBO encodes the UC logic and spinning reserve penalties based on its associated forecast profile, using a consistent base penalty structure adapted to scenario-specific parameters. Penalty terms not related to the power balance constraint $\mathcal{P}^*$ are compiled to form the base Hamiltonian $H^*$.

**Conquer (Recombination) Phase.** In the conquer phase, the algorithm aggregates sub-Hamiltonians by estimating the expectation cost using classical Monte Carlo or QCMC simulation. Rather than merging QUBOs into a single large model, the approach aggregates solution statistics after independent optimizations, thereby avoiding exponential growth in computational complexity. The strategy approximates the full stochastic problem without requiring simultaneous optimization over the entire scenario space, thereby maintaining tractability while preserving fidelity in cost estimation. For each uncertainty sample $\mathcal{u}$, which includes demand fluctuations, wind power generation, and solar PV power generation, the power balance penalty (15a) is compiled as a sub-Hamiltonian $H'_{\mathcal{u}}$. New Hamiltonians incorporating uncertainty are constructed as $\mathcal{H}$, derived from the sum of the base Hamiltonian $H^*$ and the sub-Hamiltonian $H'_{\mathcal{u}}$. The proposed divide-and-conquer method is implemented as a non-iterative batch process, enabling modular compilation and scalable optimization of QUBO Hamiltonians across large uncertainty sets. The complete set of Hamiltonians for all uncertainty samples $\mathcal{U}$ ais represented as $H'$. Computation of the cost expectation $\langle F^* \rangle$ can be computed expanding from (6b) as (16a) – (16d):

$$F^* = \{\min H' \;\; \forall \mathcal{u} \in \mathcal{U}\} \quad (16a)$$

$$\langle F^* \rangle = MC_q(\text{prob}(F^*), \mathcal{u}) \quad (16b)$$

$$\mathcal{R}|F\rangle|0\rangle = \sqrt{1 - v(F)}|0\rangle + \sqrt{v(F)}\,|1\rangle \quad (16c)$$

$$\mathcal{A} = \sqrt{1 - \text{prob}(F^*)}|0\rangle + \sqrt{\text{prob}(F^*)}\,|1\rangle \quad (16d)$$

Stochastic optimization is then conducted by calculating deterministic UC costs $F^*$ (16a) for the uncertainty set $\mathcal{U}$. Cost expectations $\langle F^* \rangle$ are estimated using QCMC $MC_q$. $MC_q$ expects a probability function $\wp$ and a problem characteristic function $v(x)$ as defined in (4b) and (5). For stochastic optimization, $\wp$ represents the probabilities of stochastic unit commitment costs $\text{prob}(F^*)$ as (16b), and the representative function is the general cost function $F$ from (7). The choice between classical MC and QCMC involves an explicit tradeoff. Classical MC computes $\langle F^* \rangle$ by directly averaging over $F^*$, requiring oracle queries proportional to $|\mathcal{U}|$. QCMC encodes $\text{prob}(F^*)$ into the amplitude of a quantum state via unitary $A$ in (16d), applies the Grover operator $Q$, and extracts $\langle F^* \rangle$ via phase estimation (6c) achieving a quadratic reduction to $O\left(\sqrt{|\mathcal{U}|}\right)$oracle queries. Equation (16) is valid because each $F_{\mathcal{u}}^*$is an independent Stage 2 dispatch cost conditioned on the fixed Stage 1 commitment schedule, forming a valid empirical distribution $\text{prob}(F^*)$ that can be encoded into $A$. The rotation $R$ in (16c) maps each cost $F$to a real-valued function on the ancilla qubit, such that the expectation recovered via phase estimation corresponds to $\langle F^* \rangle$, directly linking Phase 3 and Phase 4 of Algorithm 1.

**Algorithm 1:** Divide-and-Conquer Hybrid Quantum SUC.

**Input:** Generation Parameters $P_i^{min}, P_i^{max}, A_i, B_i, C_i, ru_i, u_m^i, d_m^i$; Uncertainty set: $\mathcal{U}$; Lagrange multipliers: $\lambda_a$, $\lambda_b$, $\lambda_c$
**Output:** Expected UC cost $\langle F^* \rangle$
*// Phase 1 – Base Hamiltonian Compilation*
Construct $H_{obj} = H_{obj}^{lin} + H_{obj}^{quad}$from (30a)–(30b)
Construct penalty terms $\mathcal{P}_2, \dots, \mathcal{P}_{11}$from (33)–(42) using $\lambda_b$, $\lambda_c$
Compile $H^* = y^T\left[H_{obj} + \sum_{k \in K, k \neq 2} \lambda_k \mathcal{P}_k\right]y$
Store $H^*$
*// Phase 2 – Scenario Sub-hamiltonian Injection*
**for each** $\mathcal{u} \in \mathcal{U}$ **do**
  Compute residual demand: $p_{u,t}^r = L_t - p_{u,t}^{wt} - p_{u,t}^{pv} \forall t \in T$ from (15b)
  Compile scenario power balance sub-Hamiltonian $H'_{\mathcal{u}}$using $\lambda_a$from (15a)
  Assemble: $H_{\mathcal{u}} = H^* + H'_{\mathcal{u}}$

**end for**
Fix commitment schedule $x_{t,i}$, $x_{t,i}^{\uparrow}$, $x_{t,i}^{\downarrow}$ from Stage 1
*// Phase 3 – Hybrid Quantum Optimization*
*// Optimize Stage-2 dispatch only via scenario $H_u$*
**for each** $u \in \mathcal{U}$ **do**
    Submit $H_u$ to solver (classical, quantum annealer, or CQM)
    Record optimal cost $F_u^* = \min H_u$
**end for**
Collect solution set: $F^* = \{F_u^* \mid \forall u \in \mathcal{U}\}$
*// Phase 4 – Cost Expectation Estimation*
**if** Classical MC **then**
    Construct cost histogram over $|\mathcal{U}|$ bins
    Compute weighted average: $\langle F^* \rangle = \sum F^* \cdot \mathrm{prob}(F^*)$
**else if** QCMC **then** *(see Fig. 4)*
    Encode $\mathrm{prob}(F^*)$ as probability unitary $A$ from (16d)
    Apply Grover operator $Q$ with rotation $R$ from (16c)
    Extract $\langle F^* \rangle$ via phase estimation (6c)
**end if**
**return** $\langle F^* \rangle$

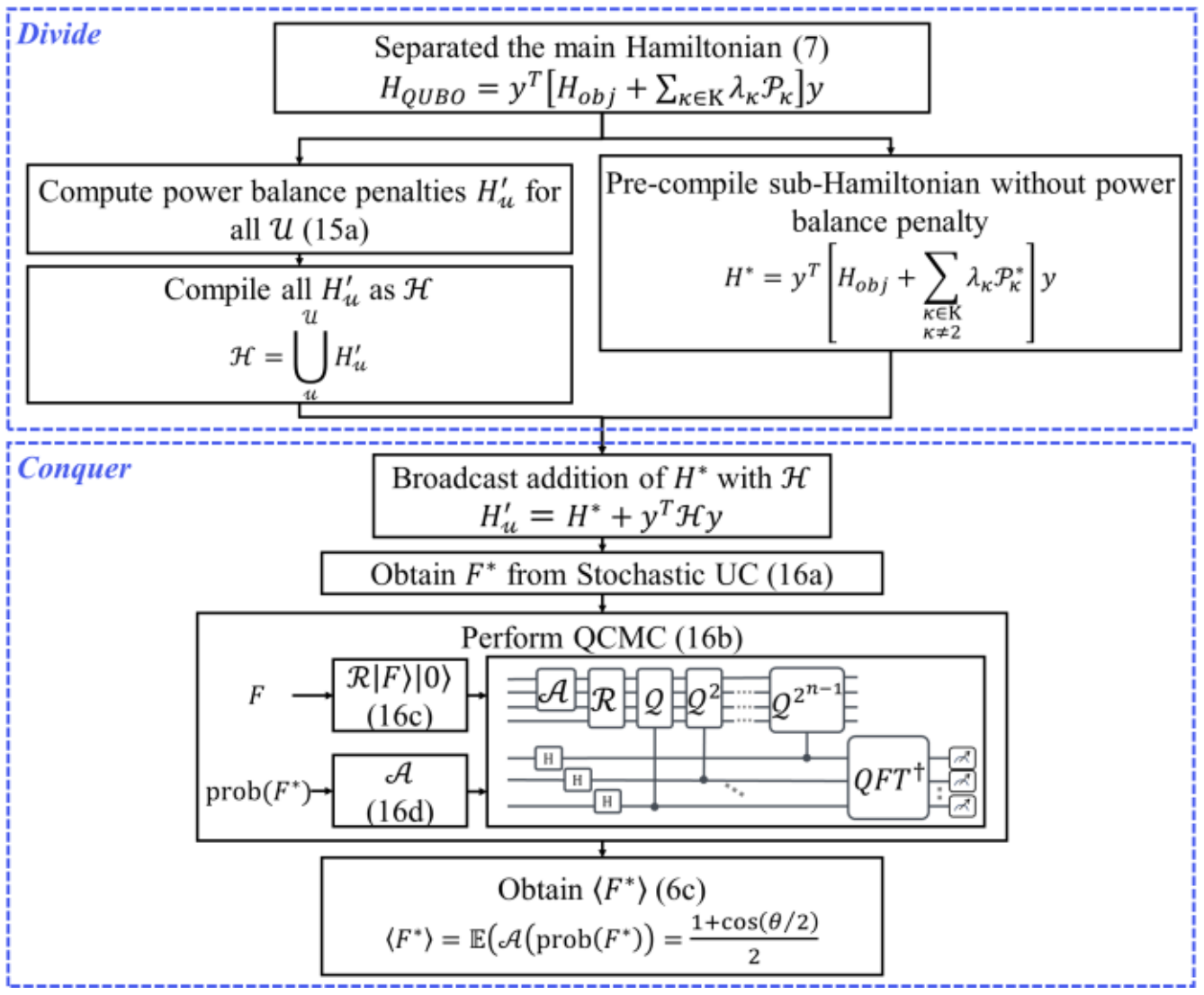

**Fig. 4** Flowchart of the divide-and-conquer algorithm for the quadratic unconstrained binary optimization in stochastic unit commitment with quantum computing Monte Carlo simulation and estimation.

## IV. Deterministic Unit Commitment Simulation Result

This section presents numerical results for deterministic UC using multiple model formulations and optimization strategies implemented within the quantum computing-ready pipeline illustrated in Fig. 3. The classical approach solves the UC problem in its MIQP form using Gurobi 11.0.3 on a CPU-based workstation (Intel i7-11700, 11th Gen, 32 GB RAM). The quantum-inspired approach reformulates the problem as a QUBO model using PyQUBO 1.4.0 in Python and applies simulated annealing with 1,240 reads on the same hardware platform. In contrast, the hybrid approach encodes the UC problem as a constrained quadratic model (CQM) using D-Wave Ocean SDK 6.x, which supports MIQP formulations. Optimization is performed via a hybrid framework that leverages CPUs, GPUs, and QPUs on the D-Wave Advantage 4.1 system, with a time limit of 5,000 ms. For QUBO-based quantum annealing, the Embedding Composite sampler is employed with a chain strength of 12,000 and up to 1,240 samples. QCMC circuits are simulated using PennyLane 0.38.0 with a CPU-based statevector backend and 4,096 shots. Scenario generation is performed using NumPy 1.26.4, yielding 27,730 independent samples, as determined by (14a). It is noted that all reported optimization times exclude embedding time for quantum annealing and queue latency associated with the hybrid CQM solver.

Scalability is evaluated across three benchmark systems: the IEEE 6-bus system [25] (3 generators, 6-hour horizon), an 11-generator system [27] (24-hour horizon), and the IEEE 118-bus system [28] (54 generators, 24-hour horizon). The IEEE 118-bus system represents a portion of the 1960s American Electric Power System, featuring 54 generators, 118 buses, and 186 transmission lines, with smaller benchmarks used for validation before scaling. Scenario generation uses a fixed random seed of 333 throughout all experiments to ensure reproducibility, with uncertainty modeled via the Beta, Weibull, and normal distributions as defined in (11)–(13). The implementation code will be made publicly available upon publication.

### A. Objective Values and Optimization Performance

Objective values and generation schedules are evaluated to assess the accuracy of hybrid quantum-classical UC optimization. Table II compares classical, quantum-inspired, and quantum methods using pure and hybrid solvers, using the classical MIQP formulation as the baseline. The quantum-inspired approach, which applies simulated annealing to QUBO models on a CPU workstation, yields objective values of \$44,719.80, \$682,089.37, and \$1,893,060.46 for the IEEE 6-bus, 11-generator, and 118-bus systems, respectively. Furthermore, the Quantum Annealing (pure quantum method) can only solve the IEEE 6-bus problem and was unable to converge for the 11-generator and IEEE 118-bus system.

The results in Table II underscore the limitations of using pure QUBO models for both classical and quantum-inspired setups. The quantum-inspired method, derived from a fourth-order PUBO reformulated into QUBO, suffers from decomposition overhead and exhibits sensitivity to externally tuned Lagrange multipliers. As system size increases, errors grow due to a reduction in complexity and the loss of constraint fidelity during PUBO-to-QUBO conversion. Table III further reveals that QUBO models executed on quantum annealers exhibit constraint violations, particularly in ramping limits, despite being labeled feasible, highlighting the challenge of encoding hard constraints via penalties alone in noisy quantum environments. In contrast, the hybrid quantum approach consistently matches the classical baseline across all systems and produces generation schedules comparable to classical solutions (Fig. 5). This is achieved by combining QUBO and MIQP formulations within a CQM framework and leveraging CPUs, GPUs, and QPUs. Constraint accuracy is ensured using Lagrange multipliers computed via D-Wave's Hybrid Quantum Solver, effectively resolving the limitations seen in quantum-inspired methods [23]. The limitations of the classical and purely QUBO models, as well as the advantages of the hybrid solver, are similar to the findings of [15].

The monotonic increase in objective error with system size from 0.22% for the 6-bus system to 2.18% for the IEEE 118-bus system reflects the cumulative effect of discretization approximation. With $N = M = 4$, each continuous power output variable is encoded using 4 binary steps; as generator count and diversity increase, representable generation values deviate further from the MIQP optimum, compounding across units and timesteps. This illustrates the fundamental accuracy-qubit tradeoff in QUBO encoding: finer discretization reduces encoding error but increases $n_{total}$ proportionally through $n_c$ and $n_s$ in (17b)-(17c), pushing qubit requirements beyond current hardware limits.

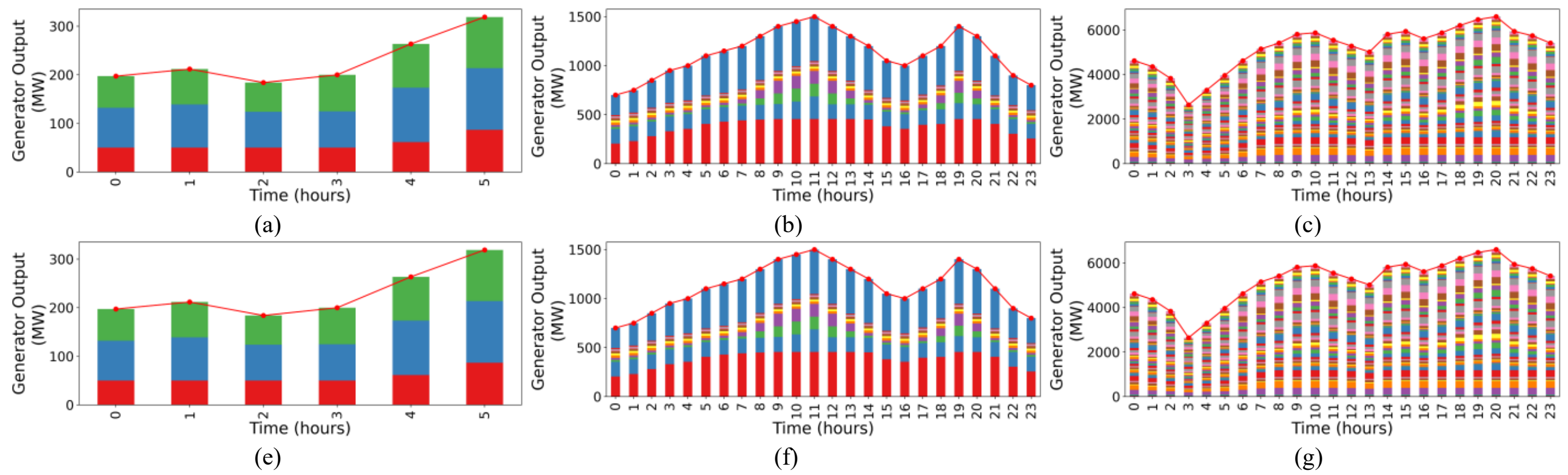

**Fig. 5** Classical unit commitment (UC) optimization using mixed-integer quadratic programming (MIQP) (a–c) and optimization using the D-Wave Hybrid Solver (e–g) solve the unit commitment problem for (a/e) the IEEE 6-bus test system, (b/f) the 11-generator benchmark system, and (c/g) the IEEE 118-bus test system. The red envelope curve represents the total power demand profile over the scheduling horizon. The colored stacked bars beneath the envelope represent the individual generator outputs at each time step.

TABLE II
PERFORMANCE COMPARISON OF DETERMINISTIC UC METHODS

| Method | Generators | Objective Value ($) | % error | time/no. of iterations |
|---|---|---|---|---|
| Classical MIQP (baseline) | 3 | 44,676.40 | - | 40.00 ms / 20 |
| | 11 | 672,084.78 | - | 130.00 ms / 513 |
| | 54 | 1,852,672.21 | - | 9,420.00 ms / 3722 |
| Quantum-inspired QUBO | 3 | 44,719.80 | 0.22% | 1.02 ms / 1024 |
| | 11 | 682,089.37 | 0.45% | 123.40 ms / 1437 |
| | 54 | 1,893,060.46 | 2.18% | 16,040.00ms /2726 |
| Quantum Annealing QUBO | 3 | 45,721.80 | 2.01% | 11.02 ms / 8192 |
| | 11 | - | - | 10,000.00 ms / - |
| | 54 | - | - | 20,000.00 ms / - |
| MIQP (hybrid solver) | 3 | 44,676.40 | 0.00% | 5,000.00 ms / 22 |
| | 11 | 672,084.78 | 0.00% | 5,000.00 ms / 17 |
| | 54 | 1,852,672.21 | 0.00% | 5,000.00 ms / 91 |

The runtime comparisons in Table II are conducted on a CPU-based workstation and report only optimization time, excluding embedding overhead. Classical MIQP and quantum-inspired methods both exhibit exponential runtime scaling with system size, evidenced by the 235× runtime increase from the 11-generator to the 118-bus system, with quantum-inspired methods additionally incurring MIQP to QUBO translation overhead that causes them to exceed classical runtimes at larger scales. While an HPC environment would reduce classical runtimes by a constant factor, it would not alter this fundamental scaling behavior. The quantum annealing advantage on the IEEE 6-bus system is hardware dependent and may diminish on HPC for such a computationally trivial problem. In contrast, the hybrid CQM maintains stable solving times across all system sizes by distributing computation across CPUs, GPUs, and QPUs in parallel, decoupling runtime from problem size growth. This architectural advantage is independent of clock speed and is expected to persist and widen as system size increases. It is noted that all reported optimization times reflect only solver execution time and exclude the time required for minor embedding on the quantum annealer and the compilation overhead for the hybrid CQM solver. For the quantum annealing case, embedding time on D-Wave Advantage depends on problem graph structure and typically ranges from seconds to minutes for problems of this scale, a non-trivial overhead that would reduce the runtime advantage observed for the IEEE 6-bus system. The hybrid CQM solver incurs additional latency from internet submission and queue time to D-Wave's cloud infrastructure.

## *B. Qubit Resource Estimation*

The logical qubit requirements for each test system are derived analytically from the formulation parameters (17a). Per generator $i$ per time step $t$, the binary variables contributing to the QUBO are grouped into commitment variables $n_b$, discretized continuous variables $n_c$, binary-encoded slack variables $n_s$, and PUBO-to-QUBO auxiliary variables $n_{aux}$. The commitment group comprises three binary variables $x_{t,i}$, $x^{\uparrow}_{t,i}$, and $x^{\downarrow}_{t,i}$ giving $n_b = 3$.

$$n_{total} = |D||T|(n_b+n_c+n_s+n_{aux}) \quad (17a)$$

$$n_c = (N+1)+(M+1) \quad (17b)$$

$$n_s = n_c + 1 \quad (17c)$$

$$\Delta n_{total} = 4|D||T| \quad (17d)$$

$$\Delta n_{full} = n_{full} - n_b \quad (17e)$$

The continuous power and reserve variables $p_{t,i}$ and $p^{sr}_{t,i}$ are each discretized into steps, respectively. Each discretized variable requires a binary-encoded slack at the same resolution, while the switching constraint introduces one additional pure binary slack producing $n_s$ qubits. Finally, auxiliary variables $n_{aux}$ per $(t,i)$ are introduced during PUBO-to-QUBO reduction of the cubic terms arising from the Boolean logic penalties. With $N = M = 4$, this gives $n_{total} = 27|D||T|$. Table III reports the resulting qubit counts broken down by variable group for all three test systems, alongside the Boolean compaction savings $\Delta n_{total}$ and the reference encodings $n_{total}$ and $n_b$ characterizing the resource gap between pure QUBO and CQM strategies. The table shows that $n_{total}$ grows from 486 qubits for the IEEE 6-bus system to 34,992 qubits for the IEEE 118-bus system, with the discretized continuous variables and their slacks accounting for the dominant share of the qubit budget. Systems beyond this region would require qubit resources exceeding current hardware limits, reinforcing the necessity of the divide-and-conquer strategy and CQM continuous variable offloading demonstrated in Table III. Logical qubit counts are reported rather than QUBO term counts, as interaction density is fully determined by $n_{total}$ given the block-diagonal sparsity structure of the UC Hamiltonian.

TABLE III
ANALYTICAL QUBO LOGICAL QUBIT RESOURCE ESTIMATION

| Variable Group | Generators 3 | 11 | 54 |
|---|---|---|---|
| $n_b$ | 54 | 792 | 3,888 |
| $n_c$ | 180 | 2,640 | 12,960 |
| $n_s$ | 198 | 2,904 | 14,256 |
| $n_{aux}$ | 54 | 792 | 3,888 |
| $n_{total}$ | **486** | **7,128** | **34,992** |
| $n_{sav}$ | 72 | 1,056 | 5,184 |
| $n_{full}$ | 558 | 8,184 | 40,176 |

Fig. 6 further illustrates the joint scaling of $n_{total}$ with respect to $|D|$ and $|T|$, with the three test systems marked at their respective operating points. The heatmap indicates that qubit demand increases with both dimensions, and that the IEEE 118-bus system already approaches the practical capacity limits of current quantum annealing hardware.

It is noted that classical optimization times remain tractable for the tested systems. The primary contribution of this proposed pipeline is to establish a quantum-compatible UC framework and to characterize the practical limits of pure quantum and hybrid implementations as hardware matures, rather than to demonstrate current runtime superiority. As shown in Table II, classical MIQP runtime grows exponentially with a 235× increase from 11 to 54 generators, while the hybrid CQM maintains architectural stability. Extrapolating this trend, systems beyond 100 generators are projected to expose significant classical scaling challenges with qubit resource projections in Table III providing analytical scalability evidence for larger systems.

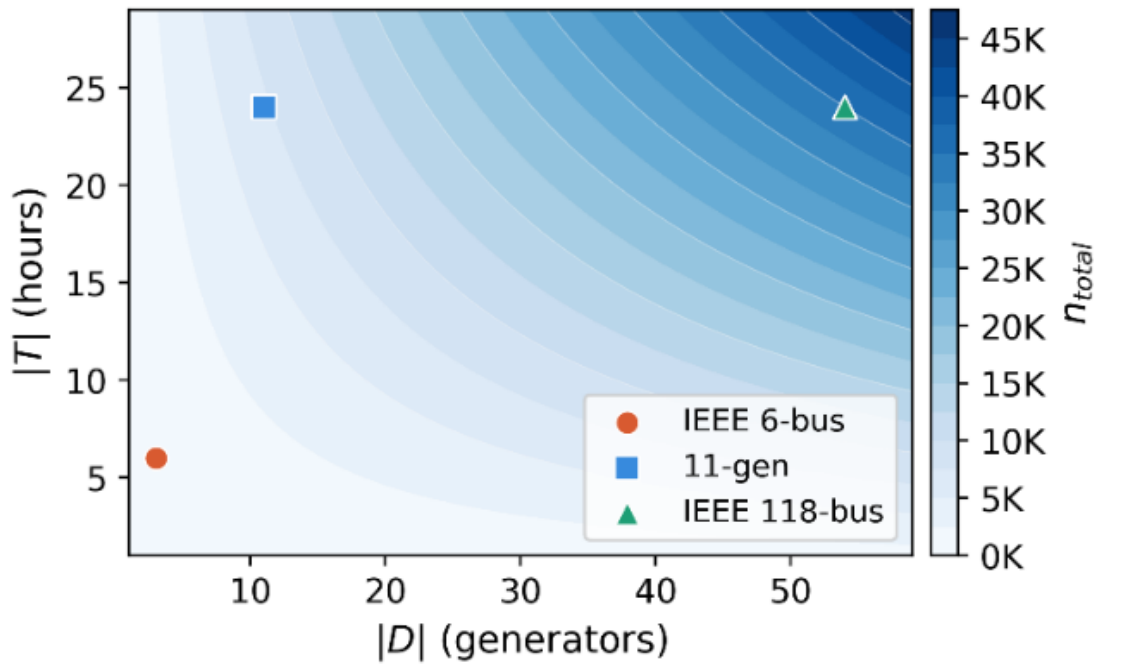

**Fig. 6** Joint scaling of the logical qubit count $n_{total}$ as a function of the number of generators $|D|$ and planning horizon $|T|$ at $N = M = 4$.

Building on the joint scaling observed in Fig. 6, the parameter-wise decomposition in Fig. 7 further reveals two distinct resource gaps relative to the pure QUBO encoding $n_{full}$. The red shaded region $\Delta n_{full}$ represents the savings from Boolean compaction, which remains a fixed proportion of $n_{full}$ independent of discretization resolution. The green shaded region $\Delta n_{total}$ represents the additional qubit reduction achieved by the hybrid CQM solver, which offloads continuous power and reserve variables to the classical solver entirely, eliminating the discretization and slack encoding overhead that constitutes the dominant share of $n_{total}$. Notably, $\Delta n_{total}$ widens with increasing $N = M$ as shown in subplot (c), indicating the CQM improvement becomes more pronounced as finer discretization is required. These trends are analytical scaling projections derived from the problem formulation, providing practical guidance for QUBO design under qubit resource constraints. This, in turn, enables informed decisions regarding discretization resolution, applicability of Boolean variable compaction, and strategies for handling continuous variables prior to solver deployment.

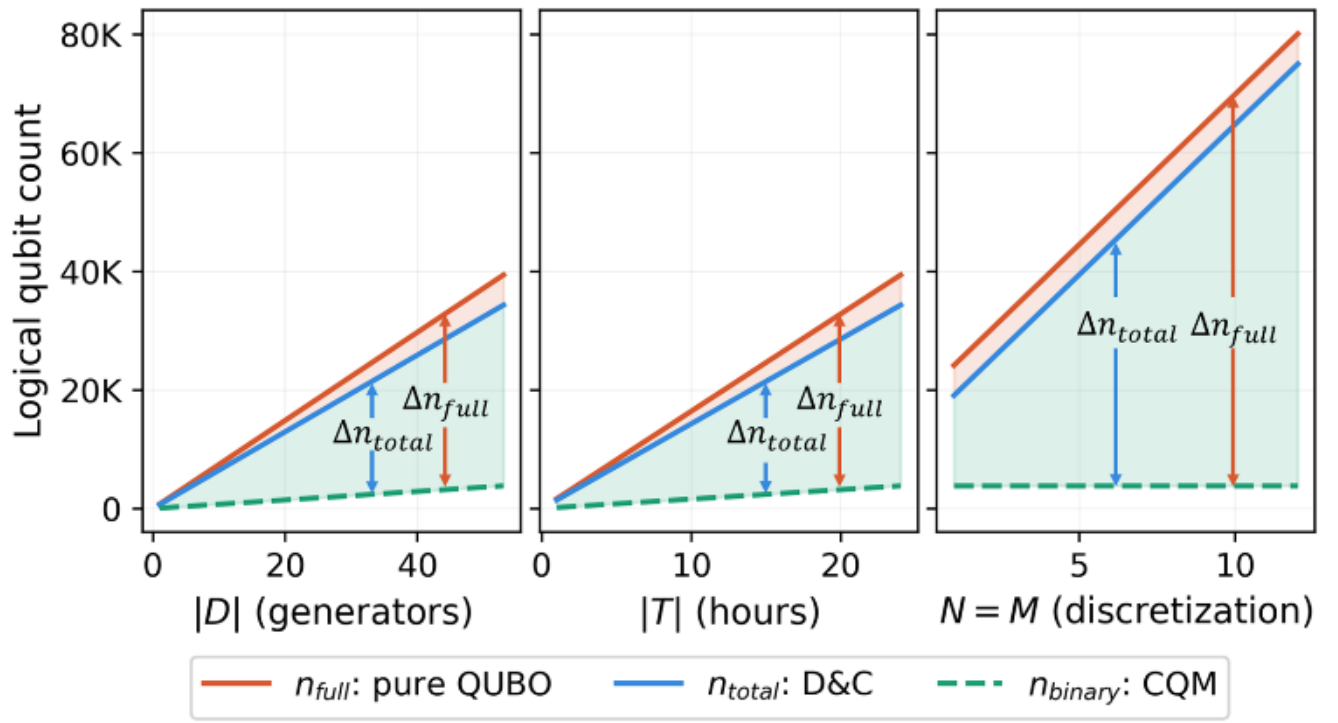

**Fig. 7** Logical qubit scaling projection for pure QUBO ($n_{full}$), divide-and-conquer ($n_{total}$), and CQM ($n_{binary}$) as a function of $|D|$, $|T|$, and discretization steps. Shaded regions indicate Boolean compaction savings $\Delta n_{full}$ (red) and CQM continuous variable offloading gap $\Delta n_{total}$ (blue).

## C. Penalty Validation

Optimization models solved using quantum-inspired, quantum annealing, or hybrid solvers are primarily evaluated based on constraint satisfaction. Ensuring constraint satisfaction or detecting penalty violations is essential for validating QUBO-based formulations. Table IV presents the satisfiability of the primary set of constraints (or penalties), which include power balance, generator start-up and cooldown, ramping, and spinning reserve requirements. The corresponding Lagrange multipliers for each penalty term are defined in (9): $\lambda_a$=22 (9a) is associated with the power balance constraint; $\lambda_b$ =30 (9b) corresponds to generator start-up, cooldown, and ramping constraints; and $\lambda_c$ =40 addresses spinning reserve requirements. The results indicate that the infeasibility observed in the quantum-inspired QUBO model for the 11- and 54-generator cases is primarily due to violations of the ramping constraints. In contrast, optimization using the hybrid solver strictly enforced all constraints, resulting in full feasibility.

TABLE IV
SUMMARY OF CONSTRAINT SATISFACTION FOR QUBO MODELS

| Method | Generators | Penalties/ Constraints: Power Balance ($\lambda_a$) | Start-up ($\lambda_b$) | Cool-down ($\lambda_b$) | Ramping ($\lambda_b$) | Spinning Reserve ($\lambda_c$) |
|---|---|---|---|---|---|---|
| Quantum-inspired QUBO | 3 | ✓ | ✓ | ✓ | ✓ | ✓ |
| | 11 | ✓ | ✓ | ✓ | ✓ | ✓ |
| | 54 | ✓ | ✓ | ✓ | ✓ | ✓ |
| Quantum Annealing QUBO | 3 | ✓ | ✓ | ✓ | ✓ | ✓ |
| | 11 | ✓ | ✓ | ✓ | × | ✓ |
| | 54 | ✓ | ✓ | ✓ | × | ✓ |
| MIQP (hybrid solver) | 3 | ✓ | ✓ | ✓ | ✓ | ✓ |
| | 11 | ✓ | ✓ | ✓ | ✓ | ✓ |
| | 54 | ✓ | ✓ | ✓ | ✓ | ✓ |

✓: constraint is satisfied × : constraint is violated

The ramping constraint violations observed for quantum annealing in Table IV stem from two distinct sources: deterministic coefficient truncation, where coupling values are clamped to D-Wave's finite precision range causing small constraint penalties to be rounded away, and stochastic environmental noise from thermal fluctuations and integrated control errors. The chain strength of 12,000 and 1,240 reads suppresses the stochastic component; the residual violations are therefore attributable primarily to coefficient truncation, which persists regardless of sample count. Potential mitigation strategies include spin-reversal transforms, penalty weight rescaling, and flux bias offset corrections. The hybrid CQM solver avoids these issues entirely by handling ramping constraints as linear inequalities on classical hardware, explaining its consistent constraint satisfaction across all systems in Table IV. An objective value sensitivity analysis considering varying Lagrangian values can be seen in Fig. 8.

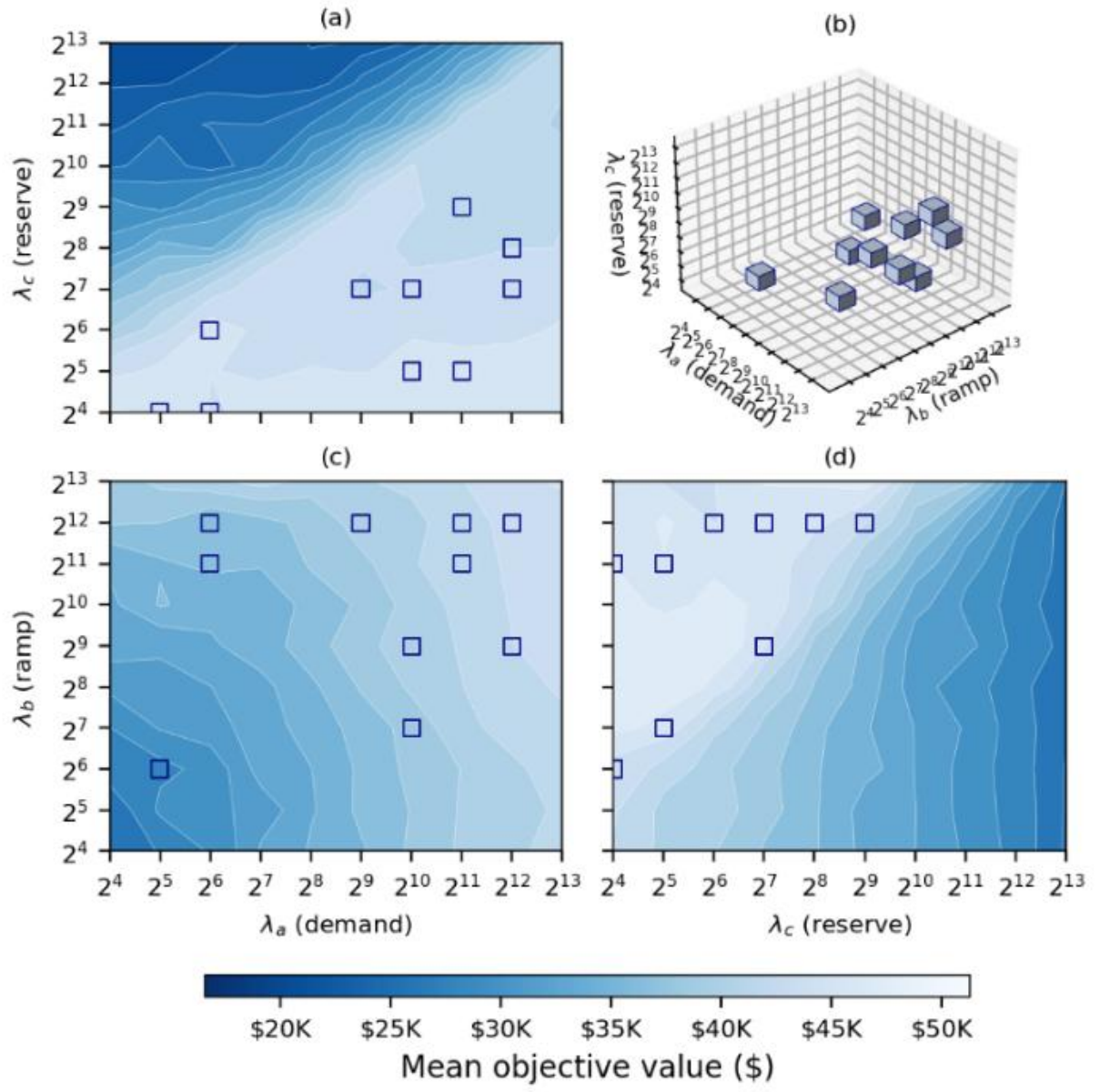

**Fig. 8** Objective value response of the quantum-inspired QUBO solver across the penalty weight space for the IEEE 6-bus system. Panels (a), (c), and (d) show filled contour maps of the mean objective value marginalized over the third penalty weight.

The Lagrange multipliers $\lambda_k$ are selected such that the penalty energy associated with any constraint violation exceeds the maximum possible reduction in objective value achievable by any single binary variable assignment which is a standard sufficiency condition for penalty-based QUBO formulations [20]. For the UC problem, $\lambda_a$ is bounded below by the maximum marginal fuel cost per unit of power imbalance, $\lambda_b$ by the maximum startup or shutdown cost differential across generators, and $\lambda_c$ by the maximum spinning reserve cost contribution. Values satisfying these bounds ensure that the QUBO energy landscape correctly prioritizes constraint satisfaction over objective minimization.

The sensitivity of solution quality to each multiplier is evaluated through a one-at-a-time analysis on the IEEE 6-bus system, presented in Fig. 8. The demand penalty $\lambda_a$ exerts the strongest influence: panels (a) and (c) show a steep objective gradient from high cost at low $\lambda_a$ to low cost at high $\lambda_a$, indicating that insufficient demand penalization leads the SA toward solutions with poor power balance. Square markers represent the combinations that recover solutions within the deterministic MIQP optimal range which are concentrated at intermediate-to-high $\lambda_a$ values, confirming a minimum penalty threshold for near-optimal convergence. The reserve penalty $\lambda_c$ produces a secondary gradient as shown in panels (a) and (d), with diminishing marginal improvement beyond a moderate threshold. The ramp penalty $\lambda_b$ shows the weakest individual effect, consistent with its smaller cost gradient relative to power balance. Panel (b) further reveals that MIQP-matching combinations cluster in a well-defined region of the joint $(\lambda_a, \lambda_b, \lambda_c)$ space, demonstrating that penalty weight selection critically governs solution quality in the QUBO formulation.

## V. Classical SUC Simulation Results

This section presents simulation results for the stochastic quantum-based UC optimization pipeline, comprising a deterministic UC baseline, uncertainty set generation, stochastic optimization, and cost expectation estimation. Performance is assessed in terms of objective quality and runtime. Cost expectations are computed using classical Monte Carlo and QCMC methods, with quantum circuits modeled and executed via the Pennylane Python package. The analysis also evaluates estimation accuracy relative to sample size and reports performance improvements from the proposed divide-and-conquer algorithm. Notably, the Hamiltonian formulation is carried out classically on a CPU, deterministic optimization is performed using either Gurobi (CPU-based) or D-Wave (quantum annealer), and QCMC sampling is simulated on a CPU using PennyLane.

### *A. SUC Classical Result Analysis*

The quality of the baseline SUC solution is evaluated in terms of its cost distribution and robustness across 27,730 uncertainty scenarios. Each scenario represents a distinct realization of demand, wind speed, and solar irradiance, and the resulting dispatch cost $F_u$ characterizes how well the non-anticipative commitment schedule performs under each uncertainty realization.

Four complementary metrics are reported. The expected cost $\langle F^* \rangle$ and standard deviation $\sigma(F^*)$ describe the central tendency and spread of the cost distribution. Value at Risk (VaR) and Conditional Value at Risk (CVaR) at the 95th percentile quantify tail exposure. The VaR marks the cost threshold exceeded by only 5% of scenarios, while CVaR captures the mean cost within that worst-5% tail. Regret, defined as the cost difference between each scenario outcome and the best-case cost, measures the suboptimality introduced by fixing commitment decisions before uncertainty is revealed. Finally, the coefficient of variation $CV = \sigma(F^*)/\langle F^* \rangle$ is computed across demand and renewable penetration quartiles to assess whether schedule performance degrades under specific operating conditions. A robust schedule is characterized by a low CV across all quartiles and a CVaR that remains close to the mean, indicating stable performance under both typical and adverse scenario combinations.

Fig. 9 presents the cost distribution and CDF of $F_u$ across all scenarios. Fig. 9(a) shows a unimodal, approximately symmetric distribution centered at $\langle F \rangle$ = \$1,813,016 with a standard deviation of \$238,714, corresponding to CV = 0.132.

The distribution exhibits a mild right skew, with a thin but nonzero tail extending toward the worst-case cost of $2,884,823, representing a 59% premium over the mean driven by simultaneous high-demand and low-renewable scenarios. The VaR at the 95th percentile is $2,215,713, meaning 95% of all scenario costs fall below this threshold, while the CVaR of $2,323,206 indicates that the average cost in the worst 5% of scenarios is approximately 28% above the mean. The mean regret of $814,131, P95 regret of $1,216,827, and maximum regret of $1,885,937 reflect the cost of non-anticipativity, representing the gap between the committed schedule and what could have been achieved with perfect foreknowledge of each scenario.

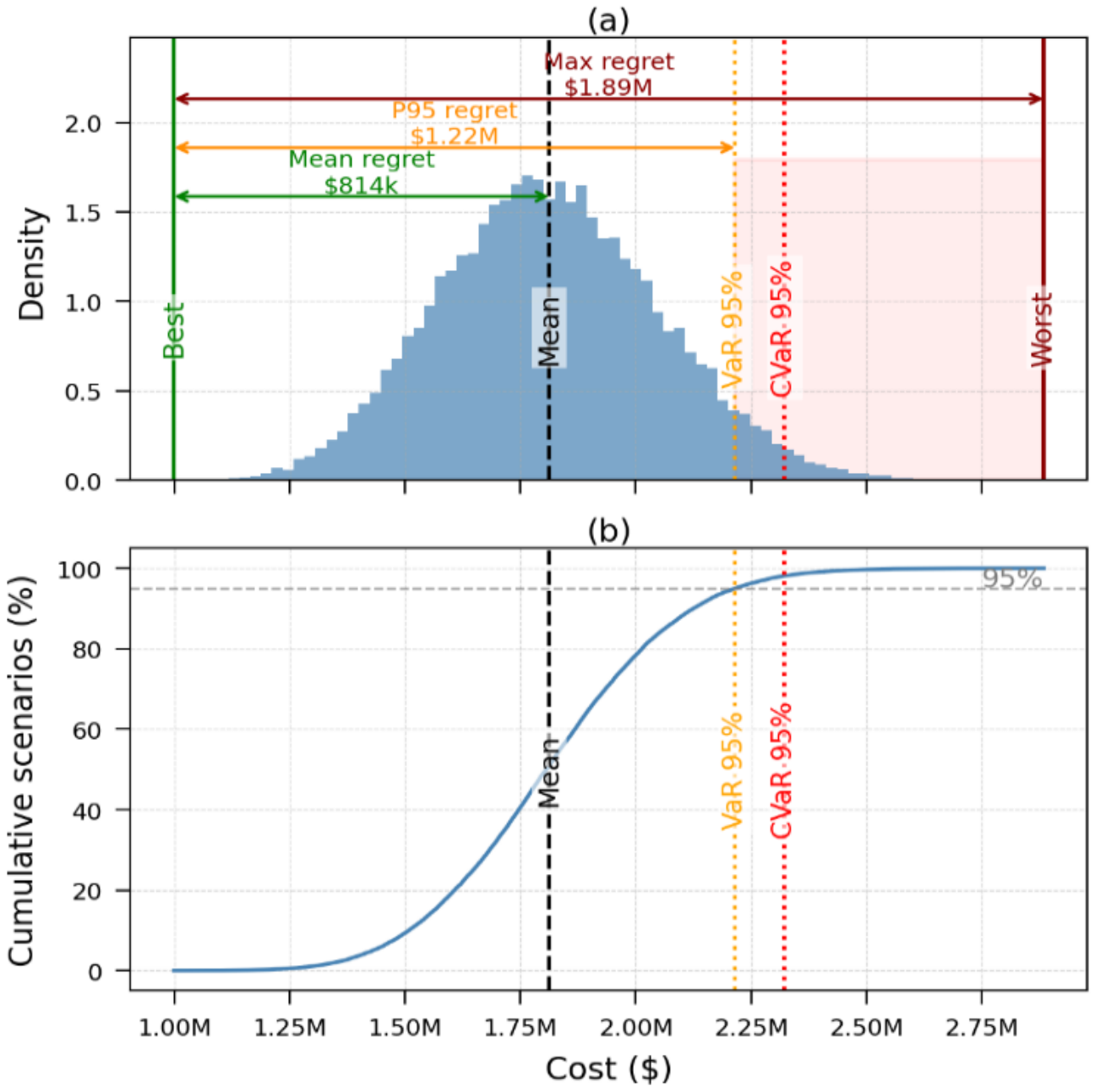

**Fig. 9** Cost distribution and CDF of the classical SUC cost. (a) the cost histogram with the expected cost, VaR and CVaR at the 95th percentile, and worst-case and best-case bounds. (b) the empirical CDF with the 95th percentile reference line.

The observations through the CDF can be confirmed in Fig. 9(b), where the steep central slope reflects the concentration of scenarios near the mean and the gradual flattening beyond the VaR threshold characterizes the tail risk region. Table V reports the CV of the dispatch cost across demand and renewable penetration quartiles. The CV remains nearly constant across all quartiles for both groupings, ranging from 0.130 to 0.133 for demand quartiles and from 0.131 to 0.133 for renewable quartiles, with ranges of 0.003 and 0.002 respectively.

TABLE V
COEFFICIENT OF VARIATION BY SCENARIO CLUSTER

| Quartile | Coefficient of Variation (CV) | |
|---|---|---|
| | Demand | Renewable |
| Q1 (Low) | 0.133 | 0.131 |
| Q2 | 0.132 | 0.133 |
| Q3 | 0.133 | 0.131 |
| Q4 (High) | 0.130 | 0.132 |
| **Range** | **0.003** | **0.002** |

The near-uniform CV indicates that the non-anticipative commitment schedule performs with consistent relative variability regardless of whether demand is low or high, or whether renewable penetration is low or high. A schedule sensitive to specific operating conditions would exhibit a monotonically increasing CV toward the high-demand or low-renewable quartiles, where generation adequacy is more constrained. The absence of such a trend confirms that the committed unit mix is well-calibrated to the full uncertainty set rather than optimized for average conditions alone, satisfying the fundamental requirement of a robust stochastic UC solution.

## *B. Comparison of Objective Values and Running Times*

The generated uncertainty set serves as an input to evaluate the stochastic UC optimization problem. The evaluation is conducted on the IEEE 118-bus system to compare the performance of classical and quantum computing approaches. The classical Monte Carlo computes the expected cost of the solution set. Table II summarizes the optimization results for an uncertainty set comprising 27,730 scenarios. Both the classical and quantum methods using the hybrid solver produce an expected cost of $1,805,658.23, consistent with the deterministic UC results reported in Section IV. In contrast, the quantum-inspired approach yields an expected cost of $1,873,189.85, corresponding to a 2.18% error relative to the classical baseline, as shown in Table II.

The total runtime for solving the stochastic UC problem is 40.76 hours using classical optimization and 14.79 hours using quantum computing and hybrid solvers, indicating significant time savings with the hybrid approach. However, the hybrid method incurs additional model compilation overhead due to internet latency and the submission process required by D-Wave's Hybrid Quantum Solvers. Compared to other approaches, the quantum-inspired method demands approximately 25.44 minutes per model and an average of 17.17 minutes for compilation, resulting in a projected total runtime of around 820 days for processing the entire uncertainty set. Fig. 10 presents the breakdown of total solution time for SUC optimization over 27,730 scenarios. Submission and network overhead are estimated based on two representative CQM solver response files, yielding an average of 0.513 s per scenario, while the QPU access time is aggregated from the same files, averaging 0.032 s per scenario. The UC optimization time is obtained as the residual after subtracting all other components. Submission, network, and QPU access components are not applicable to the classical solver.

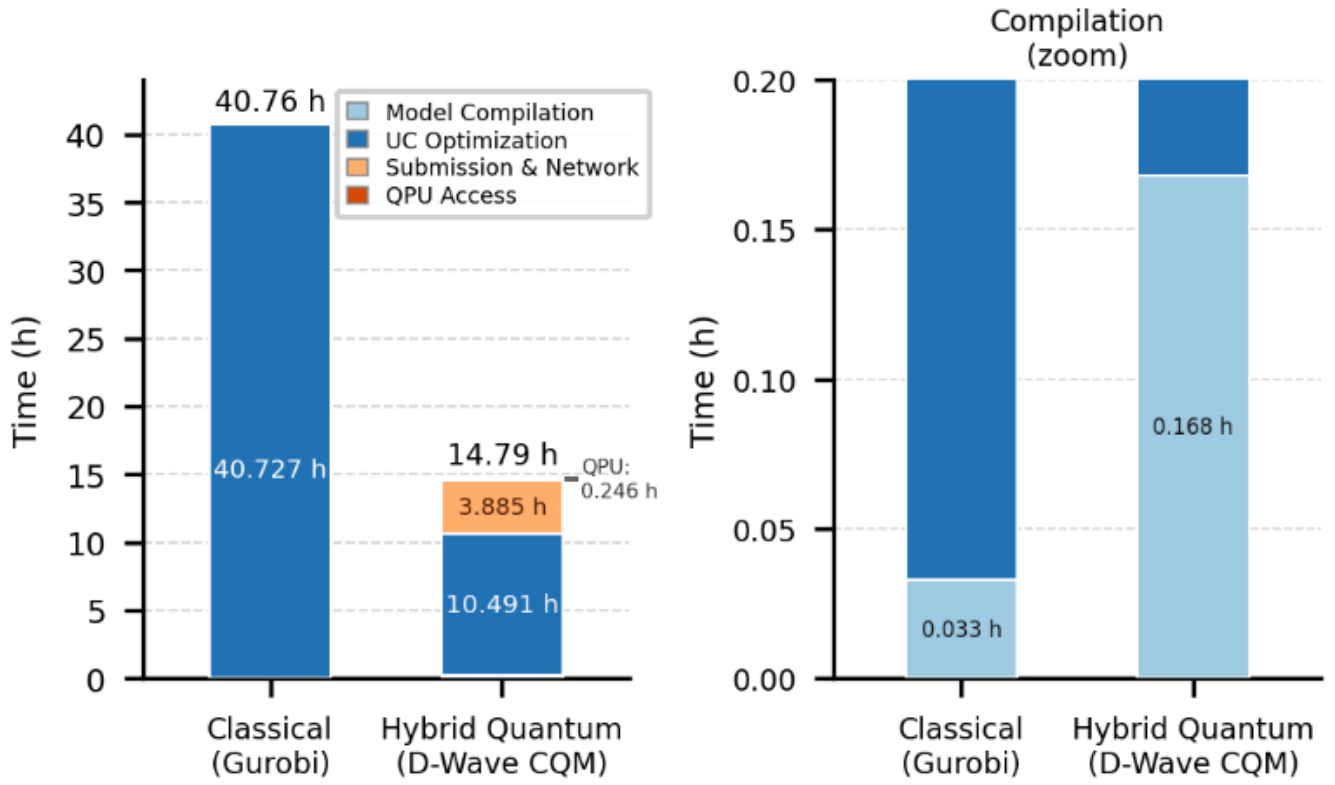

**Fig. 10** Total solution time breakdown for stochastic UC optimization over 27,730 scenarios.

Fig. 10 reports the breakdown of total solution time for SUC optimization over 27,730 scenarios, comparing the classical Gurobi solver (40.76 hours) and the hybrid D-Wave CQM solver (14.79 hours) across four components: model compilation, UC optimization, submission and network overhead, and QPU access. The submission and network overhead are estimated from representative CQM solver response files, yielding an average of 0.513 s per scenario, while the QPU access time is aggregated from the same files, averaging 0.032 s per scenario. The UC optimization time is obtained as the residual after subtracting all other components from the totals reported in Table VI.

Given the high computational cost, Table VI omits the results from the quantum-inspired and quantum-annealing QUBO models. The effectiveness of the hybrid CQM solver stems from its structural separation of the UC problem by variable type. Binary commitment variables are encoded as QUBO subproblems and submitted to the QPU, while continuous power and reserve variables are handled directly by classical hardware (CPU/GPU) without discretization [23]. This is a single-pass separation, not an iterative decomposition, requiring no subproblem exchange and no convergence criterion. Optimality is governed by classical branch-and-bound coordinating both partitions, with QPU samples informing the binary variable search [21]. This architecture explains why the hybrid solver consistently matches the classical MIQP baseline across all system sizes while maintaining a stable runtime of approximately 5,000 ms regardless of scale which is a behavior also noted in [15].

TABLE VI
COMPARISON OF EXPECTATION COSTS AND TIMES FOR STOCHASTIC UC OPTIMIZATION USING DIFFERENT METHODS

| Stochastic UC Solution Method | Classical MC Estimated Expectation ($) | Total Model Compilation Time (h) | Total Running Time (h) |
|---|---|---|---|
| Classical (Gurobi) | 1,805,658.23 | 0.033 | 40.76 |
| Hybrid Quantum Computing (D-Wave) | 1,805,658.23 | 0.168 | 14.79 |

Runtime trends in Table VI are expected to diverge further as system size increases. The classical MIQP runtime exhibits exponential growth with problem size, as illustrated by the 235× increase from the 11-generator to the IEEE 118-bus case in Table II. In contrast, the hybrid CQM approach maintains relatively stable runtimes by distributing computation across CPUs, GPUs, and QPUs. It should be noted that realizing this projected advantage at larger scales is contingent upon the availability of utility-scale or fault-tolerant quantum hardware with sufficient qubit capacity and reduced noise levels beyond those currently accessible in non-exclusive settings.

### *C. Classical MC and QCMC Simulations Cost Estimate*

Expectation cost for stochastic quantum UC can be estimated using either classical MC or QCMC methods. In both cases, estimation bins define cost intervals, with larger bin counts representing finer resolution across the search space. To align with QCMC's qubit structure, bins are expressed as powers of two, where the number of qubits corresponds to $\lfloor \log_2(|\mathcal{U}|) \rfloor$, ensuring compatibility with the uncertainty set's cardinality $|\mathcal{U}|$. The estimation bins discretize the UC cost range $[\min F^*, \max F^*]$ into $2^k$ equal-width intervals, where $k$ denotes the number of estimation qubits. Each scenario cost $F_u^*$ is assigned to its corresponding interval, forming the empirical distribution $\text{prob}(F^*)$ encoded into unitary $A$ in (16d). Bin count introduces a bias-variance tradeoff: too few bins produce coarse resolution and high bias, while too many bins spread scenarios thinly across intervals increasing estimation noise.

Table VII and Fig. 11 compare classical MC and QCMC estimates for a sample size of 27,730. QCMC achieves its most accurate result of $1,823,982.87 with a 0.60% error using 8 estimation bins. In contrast, classical MC attains an error of 0.05% (estimate: $1,805,658.23) but requires 128 bins.

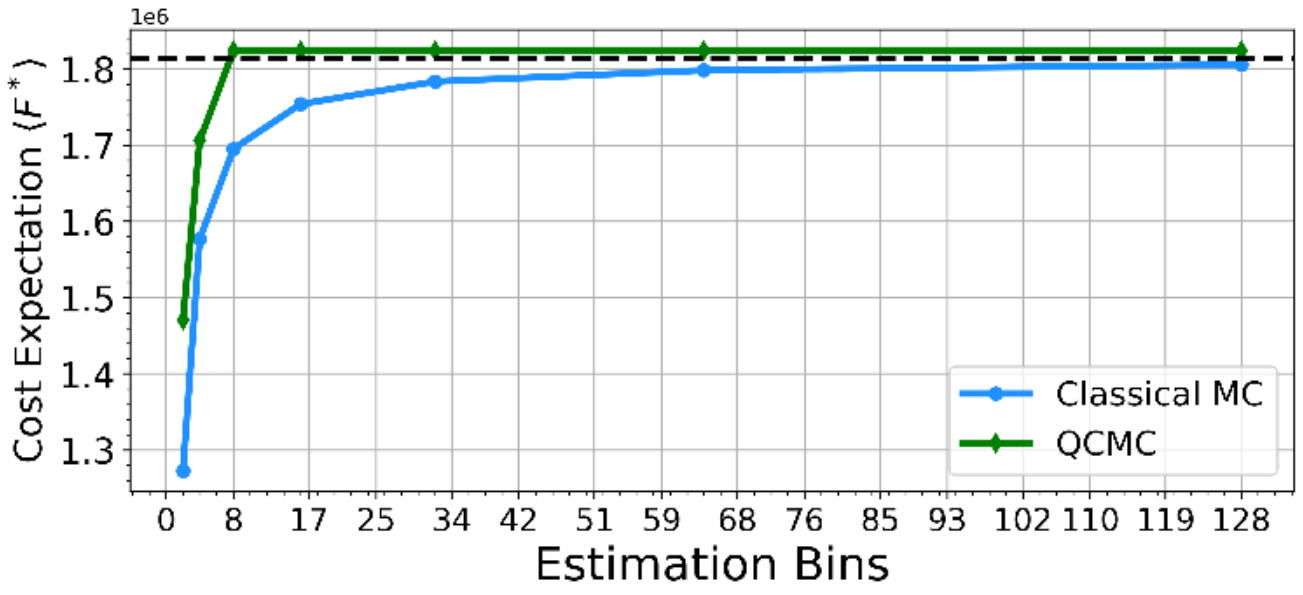

**Fig. 11** Estimated cost expectation for stochastic unit commitment (SUC) using both classical Monte Carlo (MC) and quantum computing Monte Carlo (QCMC) simulations across varying estimation bin counts. Dashed black lines represent the reference value for the expected cost.

For QCMC, each additional estimation qubit doubles the bin count while increasing the circuit depth by one layer, thereby increasing sensitivity to gate errors. The best reported QCMC accuracy of 0.60% at 8 bins (3 qubits) reflects this tradeoff between cost distribution resolution and circuit complexity. In contrast, Classical MC methods are not subject to the circuit depth penalty, explaining their accuracy of 0.05% at 128 bins. The classical MC method exhibits a computational complexity of $O(1/|\mathcal{U}|)$. QCMC, as a Grover-based amplitude estimation algorithm, achieves a quadratic speedup with complexity $O(1/\sqrt{|\mathcal{U}|})$, reducing the required number of bins from 128 to 8 while maintaining comparable accuracy. It is important to distinguish theoretical query complexity from practical execution costs. The $O(1/\sqrt{|\mathcal{U}|})$ complexity refers to the number of Grover operator queries, not wall-clock runtime as noted in Table VII. Each query involves a circuit of depth proportional to $\lfloor \log_2 |\mathcal{U}| \rfloor$ estimation qubits, and deeper circuits accumulate gate errors on real hardware, requiring additional shots to compensate.

TABLE VII
COMPARISON OF CLASSICAL AND QUANTUM COST EXPECTATION ESTIMATIONS FOR STOCHASTIC UNIT COMMITMENT

| Solving Mode | Estimation Method | $\langle F^* \rangle$ ($) | % error | Estimation bins | Complexity |
|---|---|---|---|---|---|
| Classical | Classical MC | 1,805,658.23 | 0.05% | 128 | $O(1/\|\mathcal{U}\|)$ |
| | **QCMC** | **1,823,982.87** | **0.60%** | **8** | $O(1/\sqrt{\|\mathcal{U}\|})$ |
| Hybrid-quantum | Classical MC | 1,805,658.23 | 0.05% | 128 | $O(1/\|\mathcal{U}\|)$ |
| | **QCMC** | **1,823,982.87** | **0.60%** | **8** | $O(1/\sqrt{\|\mathcal{U}\|})$ |

In this work, QCMC is simulated via PennyLane under ideal noiseless conditions, so the demonstrated advantage is the reduction in estimation bins from 128 to 8, rather than a realized runtime speedup. The full realization of the quadratic speedup requires fault-tolerant quantum hardware and is identified as a direction for future research.

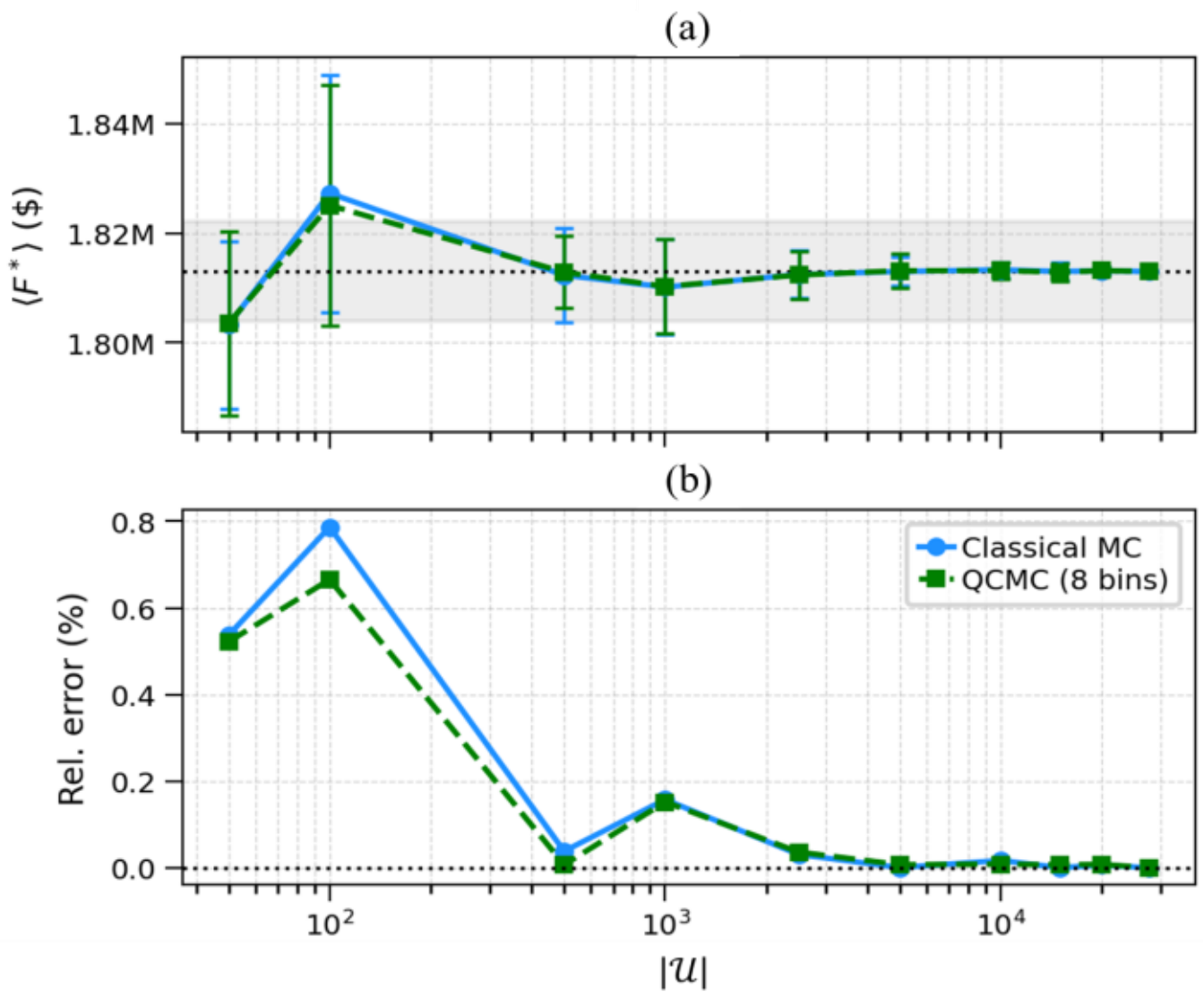

**Fig. 12** Convergence of classical MC and QCMC cost expectation estimates across increasing scenario subset sizes. (a) Estimated expected UC cost ⟨F*⟩ with standard deviation error bars and varia. (b) Relative error with respect to the full-set reference mean (n = 27,730). QCMC uses 3 estimation qubits (8 bins).

The shaded band in Fig. 12 (a) represents a ±0.5% tolerance region around the full-set reference, providing a practical convergence threshold. Fig. 12 (a) shows that both methods converge toward the full-set reference mean of $1,813,062 with error bars narrowing monotonically, confirming that estimation variance scales as $O(1/\sqrt{|\mathcal{U}|})$, consistent with the Sample Average Approximation (SAA) convergence rate. The non-monotonic spike observed at $|\mathcal{U}|$=100 in panel (b) reflects sampling variability at small subset sizes rather than a systematic effect. At such small sample sizes, random draws may over- or under-represent tail scenarios, producing temporarily elevated error before the distribution stabilizes. Table VII quantifies such behavior wherein the standard deviation decreases from approximately $15,000–$22,000 at $|\mathcal{U}|$ = 50-100 to below $800 at $|\mathcal{U}|$ = 20,000 for both methods, while relative errors fall below 0.05% beyond $|\mathcal{U}|$ = 10,000. Both methods enter the ±0.5% tolerance band at approximately $|\mathcal{U}| \approx 2{,}500$.

TABLE VIII
SCENARIO COUNT SENSITIVITY: STD. DEV. AND RELATIVE ERROR OF ⟨F*⟩ ESTIMATES

| $\lvert\mathcal{U}\rvert$ | Classical MC | | QCMC (8 bins) | |
|---|---|---|---|---|
| | Std. Dev ($) | Error (%) | Std. Dev ($) | Error (%) |
| 50 | 15,306.56 | 0.5371 | 16,839.02 | 0.5220 |
| 100 | 21,694.65 | 0.7869 | 22,011.58 | 0.6656 |
| 500 | 8,664.73 | 0.0384 | 6,600.48 | 0.0071 |
| 1,000 | 8,717.46 | 0.1575 | 8,571.51 | 0.1519 |
| 2,500 | 4,345.60 | 0.0306 | 4,337.31 | 0.0362 |
| 5,000 | 2,653.60 | 0.0005 | 3,110.28 | 0.0073 |
| 10,000 | 1,445.65 | 0.0170 | 1,545.11 | 0.0081 |
| 15,000 | 1,657.16 | 0.0003 | 1,611.52 | 0.0072 |
| 20,000 | 766.06 | 0.0062 | 798.88 | 0.0078 |

The results empirically support the scenario count of 27,730 determined by (14b). While the mean estimate stabilizes at relatively small sample sizes, the continued reduction in variance at larger $|\mathcal{U}|$ reflects improved coverage of low-probability, high-cost scenarios that are critical for robust commitment decisions under uncertainty. Notably, QCMC closely tracks classical MC across all subset sizes, exhibiting comparable standard deviations and relative errors at each point reported in Table VIII. This observation suggests that the 8-bin amplitude encoding does not introduce systematic bias relative to the continuous mean estimator over the full range of scenario counts considered. It is important to note that Fig. 11 and Fig. 12 address complementary aspects of the analysis. Specifically, Fig. 11 justifies the selection of 8 estimation bins for QCMC, whereas Fig. 12 supports the choice of 27,730 scenarios.

### *D. Performance of Divide-and-Conquer Algorithm*

This subsection evaluates the efficiency of the proposed divide-and-conquer algorithm for QUBO compilation. For large-scale QUBO models, such as those based on the IEEE 118-bus system, the average compilation time per model was approximately $t_c = 11{,}238$ seconds, with each uncertainty scenario corresponding to a distinct QUBO formulation. Consequently, the total compilation time is expected to scale linearly with the number of scenarios, given by $|\mathcal{U}| \times t_c$. Table V demonstrates the efficiency gains achieved by the proposed divide-and-conquer algorithm, compared to the naïve QUBO Hamiltonian compilation approach in which each full model is compiled sequentially and iteratively across the set $\mathcal{U}$.

The evaluation considers the IEEE 6-bus system (3 generators), the 11-generator test case, and the IEEE 118-bus system (54 generators) as representative benchmarks for small-, medium-, and large-scale UC scenarios, respectively. Due to the computational cost of compiling all 27,730 scenarios, naïve compilation times were extrapolated from 10 sample runs. Based on this extrapolation, the naïve method is estimated to require approximately 18.73 minutes for the 6-bus system, 25.23 days for the 11-generator case, and 4,097.22 days for the 118-bus system as seen in Table IX. In contrast, the divide-and-conquer algorithm reduces compilation times to 28.46, 573.75, and 1,030 seconds, respectively. These results reflect a near two-order-of-magnitude improvement, confirming the proposed method's effectiveness in reducing compilation time across all test systems.

TABLE IX
COMPARISON OF QUBO COMPILATION PERFORMANCES

| Compilation Algorithm | No. of generators | | |
|---|---|---|---|
| | 3 | 11 | 54 |
| Naïve Compilation* | $2.35\times10^3$ s | $2.18\times10^6$ s | $354.97\times10^6$ s |
| Proposed Method† | **28.46 s** | **573.75 s** | **$1.03\times10^3$ s** |

* Compilation was only done for 10 samples
† Compilation was done for all 27,730 samples

## VI. CONCLUSION AND FUTURE WORKS

This paper presents a quantum computing-ready pipeline for solving both deterministic and stochastic Unit Commitment (UC) problems in large-scale power systems. (a) The pipeline integrates classical UC modeling with quantum-compatible solvers, including simulated gate-based computing and quantum annealing, offering a unified framework for evaluating the practical advantages and limitations of quantum optimization in UC scheduling. (b) It introduces a novel QUBO reformulation that incorporates Boolean logic-based binary penalties to model startup, cooldown, and spinning reserve constraints, enabling scalability across both temporal and generator dimensions. (c) To address the complexity of stochastic UC, the study applies a divide-and-conquer strategy for efficient QUBO compilation and employs Quantum Computing Monte Carlo (QCMC) to perform sample-efficient cost estimation across uncertainty scenarios. Collectively, these contributions enhance hybrid quantum optimization approaches in power systems by improving scalability, constraint representation, and computational efficiency.

Experimental results validate the effectiveness of the proposed optimization pipeline for modern UC applications. In the deterministic UC setting, the pipeline consistently matches the optimal solutions produced by classical MIQP solvers, achieving a stable runtime of approximately 5,000 milliseconds across varying system sizes. This performance surpasses both classical heuristic approaches and quantum-inspired methods. In the stochastic case, the pipeline attains the same expected cost as the classical benchmark on the IEEE 118-bus system while reducing the runtime significantly from 40.76 to 14.79 hours. Quantum-inspired solvers were found to be impractical at scale due to excessive computation times. By integrating QCMC with a divide-and-conquer strategy, the pipeline enables efficient cost estimation and QUBO compilation across thousands of uncertainty scenarios, thereby demonstrating a realistic, quantum-ready implementation of SUC optimization, with network-constrained extensions identified as a direct path toward real-world deployment.

Looking ahead, the projected performance gains depend on two key hardware trajectories: qubit count scaling and hybrid solver scaling. For qubit scaling, current D-Wave Advantage hardware supports up to 13 generators at $N$=$M$=4 over a 24-hour horizon; this capacity is projected to increase to approximately 24 and 245 generators at 10,000 and 100,000 qubits, respectively. For hybrid solver scaling, projections indicate potential runtime improvements on the order of 100× to 1000× over classical MIQP at system sizes exceeding 200 generators, based on the observed exponential scaling of classical methods. These projected advantages remain contingent upon the availability of utility-scale or fault-tolerant quantum hardware. The results also indicate that pure quantum annealing is currently limited to small-scale systems. Encoding errors arising from discretization manifest as objective deviations ranging from 0.22% to 2.18%, along with ramping constraint violations. This limitation represents a fundamental challenge that may be mitigated through increased qubit capacity or hybrid decomposition strategies that offload continuous variables to classical solvers.

Future work may explore several directions, including the development of noise-resilient quantum hardware, industrial integration of quantum optimization frameworks, the application of quantum machine learning and robust optimization for rolling UC, and the incorporation of DC power flow constraints through Benders decomposition. In such a framework, the quantum pipeline can address the generator commitment master problem, while classical solvers handle the network feasibility subproblems. For distributed multi-area systems, the alternating ADMM provides a complementary extension path. These directions may further advance hybrid quantum approaches for SUC toward practical and scalable deployment.

## APPENDICES

### A. Classical Unit Commitment Formulations

The QUBO reformulation presented in this appendix abstracts away the transmission network. Bus-level variables, line constraints, and DC/AC power flow equations are not included. The focus is on generator scheduling, power balance, ramping, and spinning reserve constraints. The transmission network abstraction applied here is consistent with the copperplate network assumption discussed in Section III, where the rationale and limitations of this simplification are detailed.

**Objective Function.** The objective of the UC problem is to minimize the cost of power generation.

$$\min \sum_{t\in T}\sum_{i\in D} A_i p_{t,i}^2 + B_i p_{t,i} + C_i x_{t,i} \qquad (18)$$

**Physical and Security Constraints.** A typical UC problem's constraints comprise the power balance (19), minimum and maximum generation (18), minimum uptime/downtime constraints (21)-(24), ramping (25)-(26), and spinning reserve constraints (27)-(29).

$$\textstyle\sum_{i\in D} p_{t,i} = L_t \qquad \forall t\in T \qquad (19)$$

$$P_i^{max} x_{t,i} \geq p_{t,i} \geq P_i^{min} x_{t,i} \qquad \forall\, t\in T, i\in D \qquad (20)$$

$$x_{t,i}^{\uparrow} + x_{t,i}^{\downarrow} \leq 1 \qquad \forall\, t\in T, i\in D \qquad (21)$$

$$x_{t,i} - x_{t-1,i}^{down\downarrow} \geq 0 \qquad \forall\, t\in T, i\in D \qquad (22)$$

$$1 \geq x_{t,i} - x_{t-1,i} + x_{t,i}^{\uparrow} \geq 0 \qquad \forall\, t\in T, i\in D \qquad (23)$$

$$1 \geq x_{t-1,i} - x_{t,i} + x_{t,i}^{\downarrow} \geq 0 \qquad \forall\, t\in T, i\in D \qquad (24)$$

$$\textstyle\sum_{\tau=1}^{u_m^i-1} x_{t+\tau,i} \geq (u_m^i - 1)\cdot x_{t,i}^{\uparrow} \qquad \forall\, t\in T, i\in D \qquad (25)$$

$$\textstyle\sum_{\tau=1}^{d_m^i-1} 1 - x_{t+\tau,i} \geq (d_m^i - 1)\cdot x_{t,i}^{\downarrow} \qquad \forall\, t\in T, i\in D \qquad (26)$$

$$x_{t,i} P_i^{max} \geq p_{t,i}^{sr} + p_{t,i} \geq 0 \qquad \forall\, t\in T, i\in D \qquad (27)$$

$$p_{t,i}^{sr} = SR \textstyle\sum_{i\in D} x_{t,i} P_i^{max} \qquad \forall t\in T \qquad (28)$$

$$x_{t,i} ru_i \geq p_{t,i}^{sr} \geq 0 \qquad \forall\, t\in T, i\in D \qquad (29)$$

### B. Quantum Unit Commitment Formulations

The classical MIQP UC problem reformulation to a QUBO formulation is shown herein. A quadratic cost function for fuel cost is expressed in (30), and penalties on power balance (32), switching (33), uptime and downtime (34)-(37), ramping constraints (38)-(39), and spinning reserve (40)-(42).

$$H_{lin} = \sum_{t\in T}\sum_{i\in D}\sum_{n=0}^{N} (A_i + B_i p_{t,i,n} + C_i {p_{t,i,n}}^2) z_{t,i,n} \qquad (30a)$$

$$H_{quad} = \sum_{t\in T}\sum_{i\in D}\sum_{n=0}^{N}\sum_{m=0}^{N} (2C_i p_{t,i,n} p_{t,i,m}) z_{t,i,n} z_{t,i,m} \qquad (30b)$$

where $z_{t,i,n}$ and $z_{t,i,m}$ are used to approximate the supposed continuous power generation variable.

$$p_{i,n} = P_{min,i} + n h_i\,, \qquad h_i = \frac{P_i^{max} - P_i^{min}}{n} \qquad (31a)$$

$$p_{i,n}^{sr} = m q_i\,, \qquad q_i = \frac{P_i^{max} - P_i^{min}}{m} \qquad (31b)$$

The power generation variable is discretized using its definition from the minimum and maximum power generation with $k$ number of steps since only binary variables are allowed in QUBO. The symbol $q$ is the discretization factor of the spinning reserve given the discretization steps $j$.

$$\mathcal{P}_1 = \textstyle\sum_{t\in T}\left(L_t - \sum_{i\in D}\sum_{n=0}^{N} p_{t,i,n} z_{t,i,n}\right)^2 \qquad (32)$$

$$\mathcal{P}_2 = \sum_{t\in T}\sum_{i\in D}\left(x_{t,i} + \sum_{n=0}^{N} z_{t,i,n} - 1\right)^2 \tag{33}$$

$$\mathcal{P}_3 = \sum_{t\in T}\left(\sum_{i\in D} x_{t-1,i}^{\uparrow} x_{t,i}\right)^2 \tag{34}$$

$$\mathcal{P}_4 = \sum_{t\in T}\left(\sum_{i\in D} x_{t-1,i}^{\downarrow} - x_{t-1,i}^{\downarrow} x_{t,i}\right)^2 \tag{35}$$

$$\mathcal{P}_5 = \sum_{t\in T}\left(\sum_{i\in D}\left(x_{t,i}^{\uparrow} \oplus x_{t,i}\right) \vee \neg\left(x_{t,i} \oplus x_{t-1,i}\right)\right)^2 \tag{36}$$

$$\mathcal{P}_6 = \sum_{t\in T}\left(\sum_{i\in D} \neg\left[\left(x_{t,i}^{\downarrow} \oplus x_{t,i}\right) \wedge \left(x_{t,i} \oplus x_{t-1,i}\right)\right]\right)^2 \tag{37}$$

$$\mathcal{P}_7 = \sum_{i\in D}\sum_{t=1}^{T-u_m^i}\left(-x_{t,i}^{\uparrow}(u_m^i - 1) + \sum_{\tau=t}^{u_m^i - 1} x_{\tau,i}\right)^2 \tag{38}$$

$$\mathcal{P}_8 = \sum_{i\in D}\sum_{t=1}^{T-d_m^i}\left(-x_{t,i}^{\downarrow}(d_m^i - 1) + \sum_{\tau=t}^{d_m^i - 1} 1 - x_{\tau,i}\right)^2 \tag{39}$$

$$\mathcal{P}_9 = \sum_{t\in T}\sum_{i\in D}\begin{pmatrix} P_i^{max} x_{t,i} - \\ \sum_n^N p_{t,i,n} z_{t,i,n} + p_{t,i,n}^{sr} z_{t,i,n} \end{pmatrix}^2 \tag{40}$$

$$\mathcal{P}_{10} = \sum_{t\in T}\left(\sum_{i\in D}\begin{pmatrix} -SR\left(P_i^{max} x_{t,i}\right)\sum_n^N p_{t,i,n} + \\ \sum_m^N p_{t,i,m}^{sr} \cdot z_{t,i,m} \end{pmatrix}\right)^2 \tag{41}$$

$$\mathcal{P}_{11} = \sum_{t\in T}\left(\sum_{i\in D} ru_i x_{t,i} - \sum_{m}^{N} p_{t,i,m}^{sr} z_{t,i,m}\right)^2 \tag{42}$$